\documentclass[aps,prd,superscriptaddress,notitlepage,showpacs,showkeys,10pt]{revtex4-1}

\usepackage{graphicx}
\usepackage{amsmath}
\usepackage{amssymb}
\usepackage{color}
\usepackage{bm}

\definecolor{purple}{rgb}{0.8,0,0.6}
\definecolor{darkgreen}{rgb}{0.00,0.6,0.00}
\begin{document}
\title{Anomalous thermoelectric phenomena in lattice models of multi-Weyl semimetals}

\author{E.~V.~Gorbar}
%\email{gorbar@bitp.kiev.ua}
\affiliation{Department of Physics, Taras Shevchenko National Kiev University, Kiev, 03680, Ukraine}
\affiliation{Bogolyubov Institute for Theoretical Physics, Kiev, 03680, Ukraine}

\author{V.~A.~Miransky}
%\email{vmiransk@uwo.ca}
\affiliation{Department of Applied Mathematics, Western University, London, Ontario, Canada N6A 5B7}

\author{I.~A.~Shovkovy}
%\email{igor.shovkovy@asu.edu}
\affiliation{College of Integrative Sciences and Arts, Arizona State University, Mesa, Arizona 85212, USA}
\affiliation{Department of Physics, Arizona State University, Tempe, Arizona 85287, USA}

\author{P.~O.~Sukhachov}
%\email{psukhach@uwo.ca}
\affiliation{Department of Applied Mathematics, Western University, London, Ontario, Canada N6A 5B7}

\begin{abstract}
The thermoelectric transport coefficients are calculated in a generic lattice model of multi-Weyl semimetals with a
broken time-reversal symmetry by using the Kubo's linear response theory. The contributions connected with
the Berry curvature-induced electromagnetic orbital and heat magnetizations are systematically taken into account.
It is shown that the thermoelectric transport is profoundly affected by the nontrivial topology of multi-Weyl
semimetals. In particular, the calculation reveals a number of thermal coefficients of the topological origin
which describe the anomalous Nernst and thermal Hall effects in the absence of background magnetic fields.
Similarly to the anomalous Hall effect, all anomalous thermoelectric coefficients are proportional to the
integer topological charge of the Weyl nodes. The dependence of the thermoelectric coefficients on the
chemical potential and temperature is also studied.
\end{abstract}

%\pacs{...}

\maketitle

\section{Introduction}
\label{sec:introduction}

Weyl materials realize a topologically nontrivial matter with low-energy electron excitations
described by gapless chiral fermions. (For recent reviews, see
Refs.~\cite{Yan-Felser:2017-Rev,Hasan-Huang:2017-Rev,Armitage-Vishwanath:2017-Rev}.)
The nontrivial topology of Weyl materials is directly related to the Weyl nodes which act as sources
of the Berry curvature in the reciprocal space \cite{Berry:1984} with the corresponding topological
charges ($n=1,2,3$) determined by the crystallographic point symmetries
\cite{Xu-Fang:2011,Fang-Bernevig:2012,Huang}. Because of the chiral nature of low-energy excitations,
the Weyl materials allow for the realization of the chiral anomaly \cite{ABJ} in condensed
matter physics. One of its direct observable consequences is a negative magnetoresistance
\cite{Aji:2012,SonSpivak,Gorbar:2013dha,Burkov:2015} which was observed experimentally in
Refs.~\cite{Li-Wang:2015,Li-He:2015,Li-Kharzeev,Ong,Hasan}.

The materials with the topological charges of the Weyl nodes greater than one are generically called
multi-Weyl semimetals. The double-Weyl ($n=2$) and triple-Weyl ($n=3$) semimetals have the quadratic
and cubic energy dispersion relations, respectively. (Note that only the Weyl nodes with topological charges less than or equal
to $3$ are permitted by the crystallographic point symmetries \cite{Fang-Bernevig:2012}.)  By using the
first-principles calculations, it was suggested that the double-Weyl nodes are realized in
$\mbox{HgCr}_2\mbox{Se}_4$ \cite{Xu-Fang:2011,Fang-Bernevig:2012} and $\mbox{SrSi}_2$
\cite{Huang}. While the usual Weyl semimetals ($n=1$) can be viewed as three-dimensional (3D) analogues of graphene,
the double- and triple-Weyl semimetals can be considered as 3D counterparts of bilayer \cite{McCann:2006}
and ABC-stacked trilayer \cite{Guinea:2006,Min-MacDonald:2006} graphene, respectively.

A widely used method for studying the transport properties of Weyl semimetals is the chiral kinetic
theory \cite{Son:2012wh,Stephanov,Son:2012zy}. The latter takes into account the Berry curvature
effects and correctly describes the chiral anomaly in parallel electric and magnetic fields. Unfortunately,
it also implies a local nonconservation of the electric charge when both electromagnetic
and strain-induced pseudoelectromagnetic fields are present. This nonconservation in the
chiral kinetic theory can be fixed by adding the Bardeen--Zumino (or, equivalently, Chern--Simons)
term in the definition of the current \cite{Gorbar:2016ygi}. The corresponding term is essentially the
same \cite{Bardeen} as in relativistic quantum field theory which defines the consistent
anomaly. (For an instructive discussion of the Bardeen--Zumino current in the context of Weyl
semimetals, see Refs.~\cite{Landsteiner:2013sja,Landsteiner:2016}.)

In the four-vector notation, the Bardeen--Zumino current reads
$j^{\mu}_{\text{{\tiny BZ}}} = -e^2\epsilon^{\mu \nu \alpha \beta} b_{\nu} F_{\alpha \beta}/(4\pi^2)$
\cite{Bardeen,Landsteiner:2013sja,Landsteiner:2016}, where the chiral shift four-vector is $b_{\nu}=(b_0,-\mathbf{b})$.
Here $b_0$ and $\mathbf{b}$ describe the energy and momentum-space separations between the
Weyl nodes, respectively. As it turns out, without the Bardeen--Zumino term with its explicit dependence
on $b_{\nu}$, the chiral kinetic theory cannot describe correctly the chiral magnetic effect \cite{Franz:2013,Basar:2014},
the anomalous Hall effect \cite{Ran,Burkov:2011ene,Grushin-AHE,Zyuzin,Goswami,1705.04576}, and
even some collective excitations \cite{Gorbar:2016ygi} in Weyl materials.

The principal difference between the realization of the chiral anomaly in high energy physics and
Weyl semimetals is the absence of ultraviolet divergences in the latter. Indeed, because of the finite
size of the Brillouin zone in lattice models, one can perform unambiguous calculations
for the electric and chiral (or, equivalently, valley) currents in the presence of background electromagnetic
and pseudoelectromagnetic fields \cite{Gorbar:2017-Bardeen,Gorbar:2017-Bardeen-II}. As expected,
the complete result includes the Bardeen--Zumino contributions.

In the case of the electric current, the Bardeen--Zumino current is universal and topologically protected in
Weyl semimetals \cite{Gorbar:2017-Bardeen} in the limit of vanishing temperature and chemical potential.
It is determined by the winding number of the mapping of
a two dimensional cross section of the Brillouin zone onto a unit sphere. The situation is different in the
case of the chiral charge and current densities. While they also contain contributions due to the chiral
Bardeen--Zumino current, the latter is not topologically protected \cite{Gorbar:2017-Bardeen-II}. In fact, it depends on
the definition of the chirality, as well as on the specific values of model parameters. While the result may
seem surprising, it stems from the fact that the concept of chirality (unlike the electric charge) is
ambiguous on the lattice.

In the present paper, we will extend our studies in
Refs.~\cite{Gorbar:2017-Bardeen,Gorbar:2017-Bardeen-II} to thermoelectric phenomena in a generic
lattice model of multi-Weyl semimetals. One of our main results will be the derivation of anomalous thermal
coefficients in a systematic way. While having a topological origin, they are not the exact analogues of the
Bardeen--Zumino term in the electric current. Largely, this is due to the fact that the corresponding currents
appear only at finite temperatures. Nevertheless, because of their explicit dependence on the chiral shift
parameter $\mathbf{b}$, these anomalous currents do resemble the Bardeen--Zumino current. In the
context of the chiral kinetic theory, for example, they also need to be added by hand.

In the literature, the thermal conductivity and thermopower of Weyl semimetals in the presence of
electromagnetic fields were investigated in Refs.~\cite{Kim:2014,Lundgren:2014hra,Spivak,Sharma:2016-Weyl}
by using a semiclassical approach of the Boltzmann equation. The corresponding approach for Weyl semimetals
is conventionally based on the linearized chiral kinetic theory. Although such a theory simplifies
calculations significantly, it is unable to naturally reproduce the topological response coefficients
proportional to the chiral shift. Even when the anomalous terms proportional
to $\mathbf{b}$ were neglected, it was shown that the chiral anomaly
plays an important role. In particular, the characteristic quadratic dependence of the thermal conductivity
on the magnetic field was predicted in the case of the temperature gradient parallel to the field.
Such a behavior is similar to the dependence of the anomalous electric conductivity on the longitudinal
magnetic field strength. However, it was also shown \cite{Kim:2014} that the magnetic field enters the
electric and thermal conductivities differently implying the breakdown of the Wiedemann--Franz law.
This was claimed to be another hallmark of the Weyl metallic phase that originates from its nontrivial
topology.

In order to describe anomalous responses, one can use the consistent chiral kinetic theory \cite{Gorbar:2016ygi}, where the Bardeen--Zumino term is added in the definition of the electric current. A more advanced way is to employ the chiral kinetic theory with the Berry curvature obtained in lattice models similarly to Ref.~\cite{Sharma:2016-Weyl}.
In the case of Weyl materials with a broken time-reversal (TR) symmetry, it was found that, in addition to
the usual magnetic-field-dependent Nernst effect, which was recently measured in NbP \cite{McCormick:2017}, there is also an anomalous Nernst response
\cite{Sharma:2016-Weyl}. Similarly to the anomalous Hall effect, the anomalous
Nernst effect is determined by a nonzero chiral shift. (It is worth noting that the effect was also predicted in Dirac semimetals  \cite{Sharma:2016-Dirac}, where the chiral shift is generated by magnetic field.) Therefore, in the framework of the kinetic
theory, it was predicted only when lattice models were employed \cite{Sharma:2016-Weyl},
but absent in linearized models of Weyl semimetals \cite{Lundgren:2014hra}.
The thermoelectric properties of
double-Weyl semimetals were studied in Ref.~\cite{Chen-Fiete:2016}, where it was shown that (i)
the transport exhibits an interesting directional dependence and (ii) the anomalous contributions
to the thermoelectric coefficients are doubled compared to the case of linearly dispersing Weyl
nodes. The anomalous Nernst and thermal Hall effects in a linearized low-energy model
of type-II Weyl semimetals \cite{Soluyanov-Type-II}, i.e., materials with a large tilt
of Weyl nodes, were investigated in Refs.~\cite{Ferreiros:2017abd,1707.06222}.

This paper is organized as follows. In Sec.~\ref{sec:model}, we introduce a generic lattice model of
multi-Weyl semimetals ($n=1,2,3$) with a broken TR symmetry and outline the key details of the formalism for studying the thermal
transport. The response to a background electric field and thermal gradient is considered in Sec.~\ref{sec:Kubo}.
The thermoelectric coefficients are calculated in Sec.~\ref{sec:Kubo-L-All}. The thermal conductivity,
the Seebeck tensor, the Wiedemann--Franz law, and the Mott relation are investigated in Sec.~\ref{sec:Kubo-thermo-all}.
The results are summarized and discussed in Sec.~\ref{sec:Summary-Discussions}.
Technical details of derivations are given in several appendixes at the end of the paper.
Throughout the paper, we use the units with $\hbar = c = 1$.

\section{Lattice model of multi-Weyl semimetals}
\label{sec:model}

Generalizing the low-energy effective Hamiltonian of a multi-Weyl semimetal with a broken TR
symmetry given in Refs.~\cite{Volovik:1988,Fang-Bernevig:2012,Li-Roy-Das-Sarma:2016},
one can find that the corresponding lattice model can be defined by the Hamiltonian,
\begin{equation}
\label{model-H-def}
\mathcal{H}_{\rm latt} = d_0 +\mathbf{d}\cdot\bm{\sigma},
\end{equation}
where $\bm{\sigma}=(\sigma_x,\sigma_y,\sigma_z)$ are the Pauli matrices and
functions $d_0$ and $\mathbf{d}$ are periodic in quasimomentum $\mathbf{k}=\left(k_x,k_y,k_z\right)$.

In the case of Weyl semimetals with the unit topological charge $n=1$, the functions $d_0$ and $\mathbf{d}$
take the following form:
\begin{eqnarray}
\label{model-d-def-be}
d_0 &=& g_0 +g_1\cos{(a_zk_z)} +g_2\left[\cos{(a_xk_x)}+\cos{(a_yk_y)}\right],\\
d_1 &=& \Lambda \sin{(a_xk_x)}, \label{model-d-def-d1}\\
d_2 &=& \Lambda \sin{(a_yk_y)},  \label{model-d-def-d2}\\
d_3 &=& t_0 +t_1\cos{(a_zk_z)} +t_2\left[\cos{(a_xk_x)}+\cos{(a_yk_y)}\right],
\label{model-d-def-ee}
\end{eqnarray}
where $a_x$, $a_y$, and $a_z$ denote the lattice spacings and the energy parameters
$g_0$, $g_1$, $g_2$, $\Lambda$, $t_0$, $t_1$, and $t_2$ are material dependent. Their characteristic
values can be obtained, for example, by fitting the dispersion relations of low-energy excitations
in $\mathrm{Na_3Bi}$. The corresponding values are given in Appendix~\ref{Sec:App-model} and are
used in our numerical calculations throughout the paper. For the sake of simplicity, below we
will assume that the lattice is cubic, i.e., $a_x=a_y=a_z=a$.

\begin{figure}[t]
\begin{center}
\hspace{-0.32\textwidth}(a)\hspace{0.32\textwidth}(b)\hspace{0.32\textwidth}(c)\\[0pt]
\includegraphics[width=0.32\textwidth]{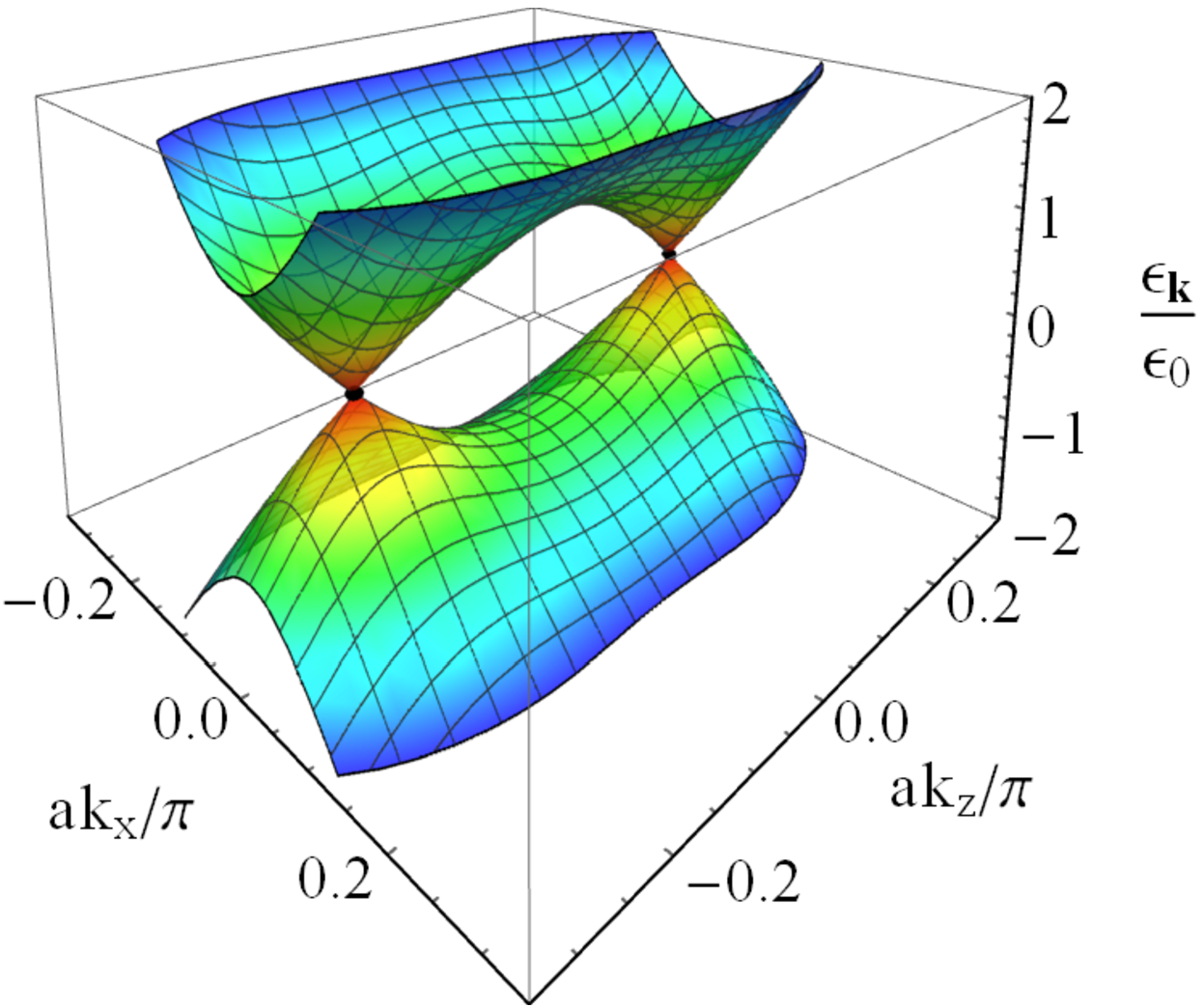}\hfill
\includegraphics[width=0.32\textwidth]{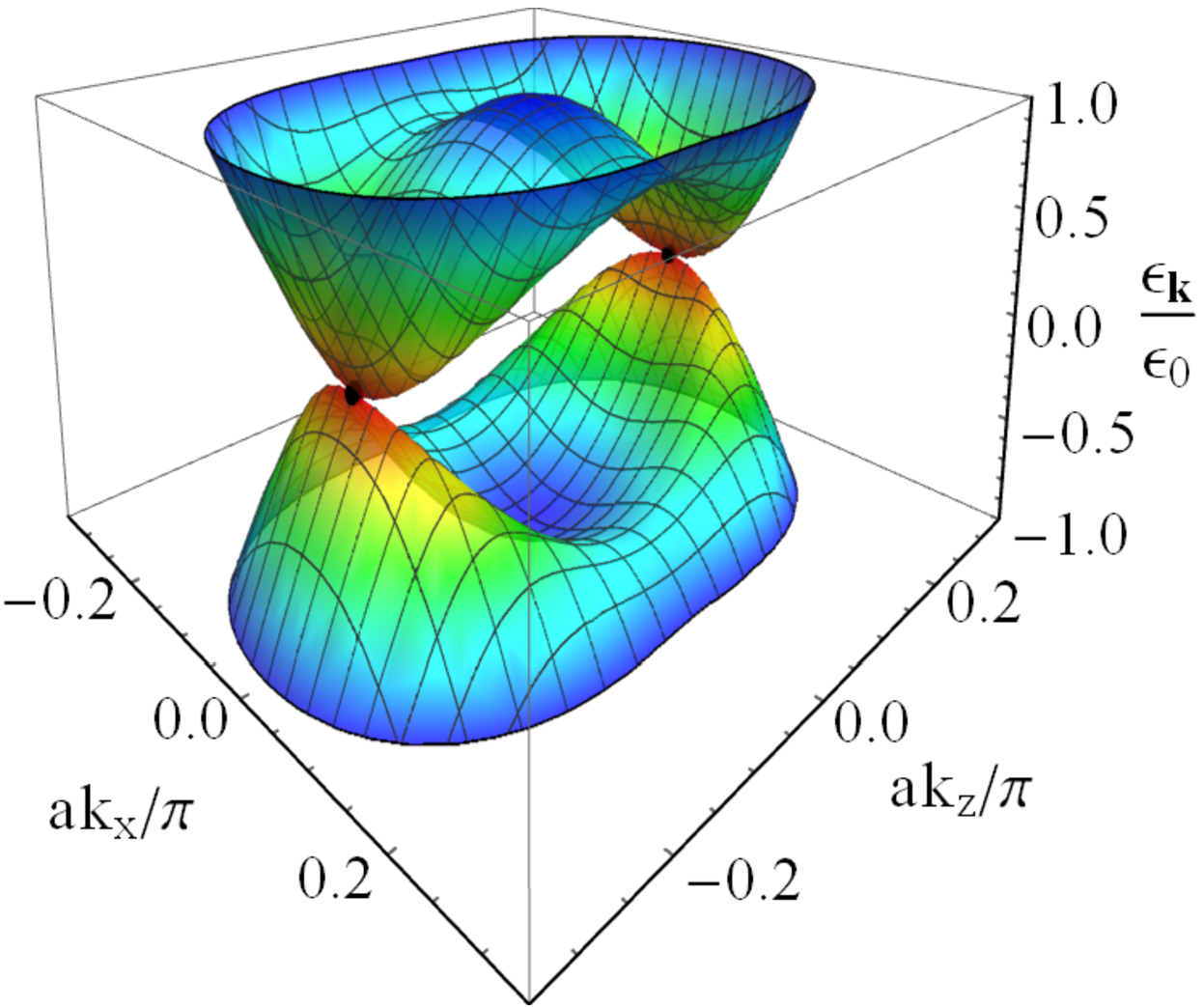}\hfill
\includegraphics[width=0.32\textwidth]{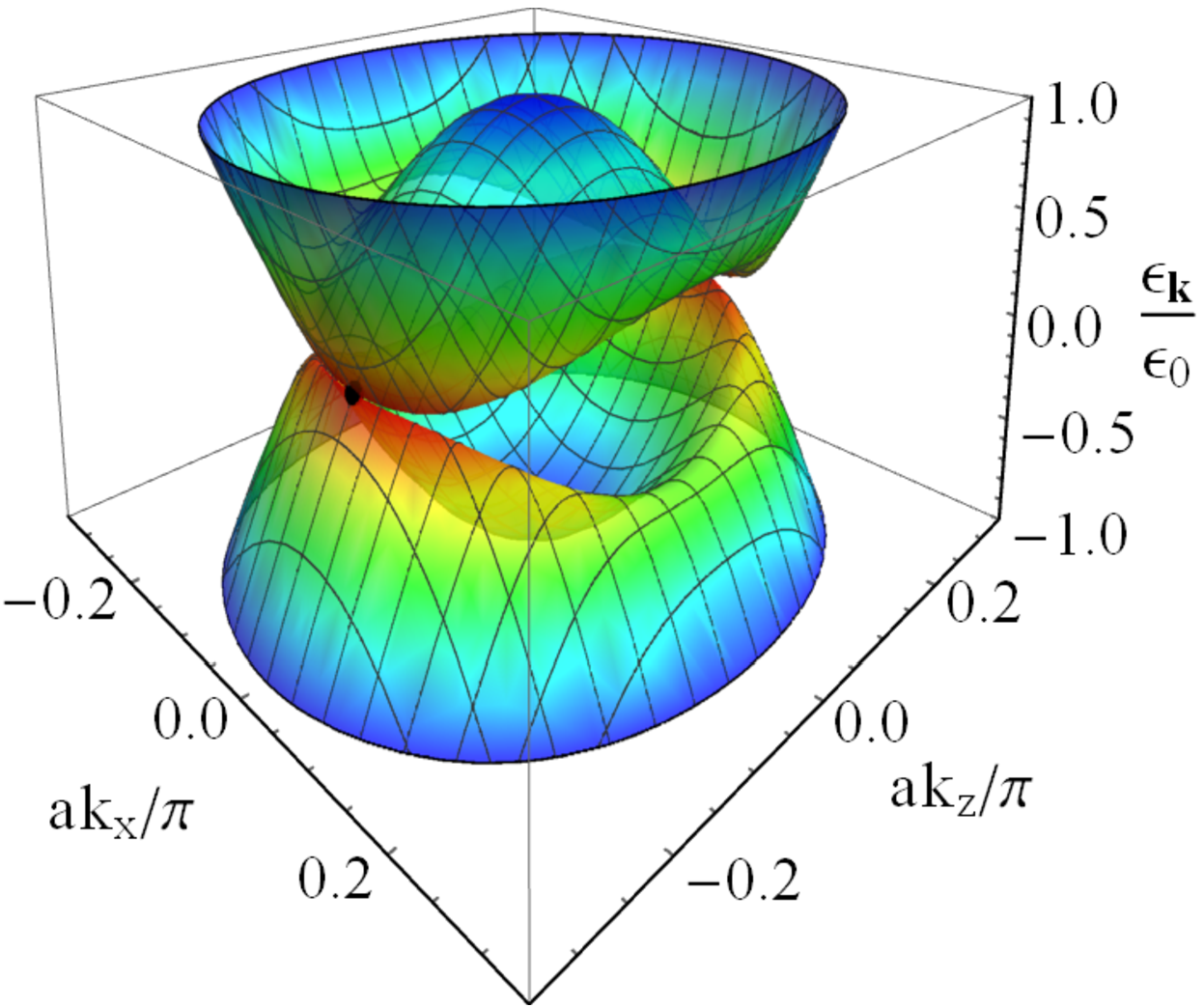}
\caption{The low-energy part of the quasiparticle spectrum in the lattice model (\ref{model-H-def})
describing multi-Weyl semimetals with the topological charges of Weyl nodes (a) $n=1$,
(b) $n=2$, and (c) $n=3$. For simplicity, we set $d_0=0$ and plotted the
energy as a function of $k_x$ and $k_z$ at fixed $k_y=0$. We also used a characteristic energy
scale set by the size of the ``dome" between the Weyl nodes $\epsilon_0\equiv |\mathbf{d}|_{\mathbf{k}=\mathbf{0}}$.
Black points label the positions of the Weyl nodes. The complete set of model parameters is given in
Appendix~\ref{Sec:App-model}.}
\label{fig:model-energy}
\end{center}
\end{figure}

For a double-Weyl semimetal with the topological charge $n=2$, one should replace
$d_1$ and $d_2$ in Eqs.~(\ref{model-d-def-d1}) and (\ref{model-d-def-d2}) with the following functions:
\begin{eqnarray}
\label{d-def-n=2-be}
d_1 &=& \Lambda \frac{\sin^2{(a_xk_x)}-\sin^2{(a_yk_y)}}{\sqrt{2}},\\
d_2 &=& \Lambda \frac{\sin{(a_xk_x)}\sin{(a_yk_y)}}{\sqrt{2}}.
\label{d-def-n=2-ee}
\end{eqnarray}
Similarly, in the case of the Weyl nodes with the topological charge $n=3$, one should use
\begin{eqnarray}
\label{d-def-n=3-be}
d_1 &=& \Lambda \frac{\sin^3{(a_xk_x)}-3\sin{(a_xk_x)}\sin^2{(a_yk_y)}}{2},\\
d_2 &=& -\Lambda \frac{\sin^3{(a_yk_y)}-3\sin{(a_yk_y)}\sin^2{(a_xk_x)}}{2}.
\label{d-def-n=3-ee}
\end{eqnarray}
As is easy to check, the dispersion relation of quasiparticles described by Hamiltonian
(\ref{model-H-def}) is given by
\begin{equation}
\epsilon_{\mathbf{k}} =  d_0 \pm |\mathbf{d}|.
\label{model-E}
\end{equation}
When the parameters are such that $|t_0+2t_2|\leq |t_1|$, this model has two Weyl nodes separated in
momentum space by distance $2b_z$, where the chiral shift
parameter $b_z$ is given by the following analytical expression:
\begin{equation}
b_z=\frac{1}{a} \arccos{\left(\frac{-t_0-2t_2}{t_1}\right)}.
\label{model-bz}
\end{equation}
For simplicity, we will assume that the quasiparticle energy vanishes at the position of Weyl nodes.
In terms of the model parameters, this implies that $g_0+2g_2-g_1(t_0+2t_2)/t_1=0$. In a general
case, this condition can be enforced by an appropriate redefinition of the reference point for the
chemical potential $\mu$. Furthermore, in order to simplify the technical details of the analysis, we will
drop the term $d_0$ altogether. While a nonzero $d_0$ introduces an asymmetry between the
valence and conduction bands, it does not affect the key topological features of the Weyl nodes and, therefore,
should not affect the main qualitative features of the thermoelectric transport. The low-energy parts
of the quasiparticle spectrum in the lattice models of multi-Weyl semimetals are presented in
Fig.~\ref{fig:model-energy}(a) for $n=1$, Fig.~\ref{fig:model-energy}(b) for $n=2$, and
Fig.~\ref{fig:model-energy}(c) for $n=3$.

%%%%%%%%%%%%%%%%%%
%%%%%%%%%%%%%%%%%%

In order to study a linear electromagnetic response, we include an interaction with the
gauge field through the usual interaction term
\begin{equation}
\mathcal{H}_{\rm int} = \mathbf{j}\cdot\mathbf{A},
\label{model-H-int}
\end{equation}
where the electric current density operator in the momentum space is given by
\begin{equation}
\mathbf{j}(\mathbf{k}) = -e\bm{\nabla}_{\mathbf{k}} \mathcal{H}_{\rm latt}=-e \sum_{i=1}^3 \sigma_i\partial_{\mathbf{k}} d_i,
\label{model-je-def}
\end{equation}
and $e$ is a fermion charge. The thermal current operator can be defined as (see, e.g., Refs.~\cite{Ferrer:2002gf,Sharapov:2003})
\begin{equation}
\label{model-jq-alt}
\mathbf{j}^{Q}(\omega;\mathbf{k})
= e^{-1}\omega\mathbf{j}(\mathbf{k}) =- \omega\sum_{i=1}^3 \sigma_i\partial_{\mathbf{k}} d_i.
\end{equation}
Here we assume that the energy of quasiparticles $\omega$ is measured from the Fermi level. In accordance with such a
convention, the Green's function in the model described by the lattice Hamiltonian (\ref{model-H-def}) is
given by
\begin{equation}
\label{model-G0-def-be}
G^{(0)}(\omega\pm i0;\mathbf{k}) =\frac{i}{2|\mathbf{d}|}\sum_{s=\pm}s
\frac{\omega +\mu+(\mathbf{d}\cdot\bm{\sigma})}{\omega +\mu-s|\mathbf{d}| \pm i0}.
\end{equation}

\section{Transport currents and magnetizations}
\label{sec:Kubo}

In this study, in order to investigate the anomalous thermoelectric response of multi-Weyl semimetals to
a background electric field and a temperature gradient, we will follow the approach of the Kubo's linear
response theory similar to that in our paper \cite{Gorbar:2017-Bardeen}, where the topological Bardeen--Zumino
contribution to the electric current density was derived in a lattice model of Weyl semimetals.

Before proceeding to the calculation of the relevant correlators in the Kubo's linear response theory, let
us recall the phenomenological expressions for the electric and heat transport current densities in
terms of the background electric field and temperature gradient (see, e.g., Ref.~\cite{Mahan-book}),
\begin{eqnarray}
\label{Kubo-Je-Jq-gen-be}
J_n &=& e^2 L_{nm}^{11} E_{m} +eL_{nm}^{12}\nabla_m\left(\frac{1}{T}\right), \\
\label{Kubo-Je-Jq-gen-ee}
J_n^Q &=& e\frac{1}{T}L_{nm}^{21}E_{m} +L_{nm}^{22}\nabla_m\left(\frac{1}{T}\right),
\end{eqnarray}
where $n$ and $m$ are the spatial indices (i.e., $x$, $y$, or $z$) and the thermodynamic forces are
defined so that the transport coefficients obey the Onsager reciprocal relation $L_{nm}^{12}=L_{mn}^{21}$.
(Note that our definition of $L_{nm}^{11}$ differs from that in Ref.~\cite{Mahan-book} by a factor of $T$.)
As is clear from Eq.~(\ref{Kubo-Je-Jq-gen-be}), the transport coefficients $L_{nm}^{11}$ and
$L_{nm}^{12}$ define the electric current densities induced by a background electric field and
temperature gradient, respectively. The coefficient $L_{nm}^{11}$ is directly related
to the electric conductivity tensor $\sigma_{nm}$, i.e., $L_{nm}^{11}\equiv \sigma_{nm}/e^2$.
From Eq.~(\ref{Kubo-Je-Jq-gen-ee}), we see that $L_{nm}^{21}$ and $L_{nm}^{22}$ define the heat current
density in response to an electric field and temperature gradient, respectively.

Let us start by reminding why the standard Kubo's formalism is unable to capture the thermoelectric
coefficients $L_{nm}^{12}$, $L_{nm}^{21}$, and $L_{nm}^{22}$ correctly in a general case. In
particular, it may fail when nonzero gradients of the chemical potential and/or temperature are present
\cite{Luttinger,Smrcka:1977}. The root of the problem is connected with the thermodynamic nature of
driving forces, which cannot be captured by an interaction Hamiltonian alone without a simultaneous
adjustment of a local (as opposed to global) thermodynamic equilibrium.

By following the Luttinger's
approach \cite{Luttinger}, it was shown in Refs.~\cite{Cooper-Ruzin:1997,Qin-Niu:2011} that there are
additional terms in the local currents that are related to the electromagnetic orbital magnetization
$\mathbf{M}$ and the so-called heat magnetization $\mathbf{M}^Q$. (The latter is a combination
of the gravitomagnetic energy and orbital magnetizations $\mathbf{M}^Q=\mathbf{M}^E-\mu\mathbf{M}/e$.)
The corresponding magnetizations are responsible for two different types of local currents. One of them
is the divergence-free current $\sim\bm{\nabla} \times\mathbf{M}$ (or $\sim\bm{\nabla}\times\mathbf{M}^{Q}$)
that circulates locally and, therefore, does not affect the net transport current flowing through the system. The
other is an additional transport current which is proportional to the thermodynamic forces and the local
magnetization \cite{Cooper-Ruzin:1997,Qin-Niu:2011}. The inclusion of the latter is essential for the
correct description of the thermoelectric response, as well as for reproducing the Onsager reciprocal relations.
According to Refs.~\cite{Cooper-Ruzin:1997,Qin-Niu:2011}, the transport coefficients $L_{nm}^{\alpha \beta}$
with $\alpha,\beta=(1,2)$ are given by the following relations:
\begin{eqnarray}
\label{Kubo-magnetization-all-repl-be}
L_{nm}^{11} &=& K_{nm}^{11},\\
L_{nm}^{12} &=& K_{nm}^{12}- \frac{T}{e}\epsilon_{nml}M_l,\\
L_{nm}^{21} &=& K_{nm}^{21}- \frac{T}{e}\epsilon_{nml}M_l,\\
L_{nm}^{22} &=& K_{nm}^{22}- 2T\epsilon_{nml}M^{Q}_l,
\label{Kubo-magnetization-all-repl-ee}
\end{eqnarray}
where $K_{nm}^{\alpha\beta}$ denote the corresponding coefficients calculated in the standard Kubo's linear response
theory and $\epsilon_{nml}$ is an antisymmetric tensor. We will derive the expressions for the coefficients $K_{nm}^{\alpha\beta}$, as well as the relevant
magnetizations $\mathbf{M}$ and $\mathbf{M}^{Q}$ in the next two subsections.

\subsection{Kubo's linear response theory}
\label{sec:Kubo-K}

In the Kubo's linear response theory the transport coefficients $K_{nm}^{\alpha\beta}$ are defined in terms of the
current-current correlation functions. By making use of the electric and heat current operators in Eqs.~(\ref{model-je-def})
and (\ref{model-jq-alt}), respectively, we derive the following general expressions for the relevant coefficients:
\begin{eqnarray}
\label{Kubo-L-11-A}
K_{nm}^{11} &=& -\frac{1}{e^2}\,\mbox{Re}\left(\lim_{\Omega\to0}\frac{i}{\Omega} T \sum_{l=-\infty}^{\infty} \int\frac{d^3\mathbf{k}}{(2\pi)^3}
\int \int d\omega d\omega^{\prime} \frac{\mbox{tr}\left[j_{n}(\mathbf{k})A(\omega; \mathbf{k})j_{m}(\mathbf{k})
A(\omega^{\prime}; \mathbf{k})\right]}{\left(i\omega_l+\mu-\omega\right)\left(i\omega_l-\Omega-i0+\mu-\omega^{\prime}\right)}\right),\\
\label{Kubo-L-12-A}
K_{nm}^{12} &=& -\frac{T}{e}\,\mbox{Re}\left(\lim_{\Omega\to0}\frac{i}{\Omega} T \sum_{l=-\infty}^{\infty} \int\frac{d^3\mathbf{k}}{(2\pi)^3}
\int \int d\omega d\omega^{\prime} \frac{\mbox{tr}\left[j_{n}(\mathbf{k})A(\omega; \mathbf{k})j_{m}^Q(\mathbf{k})
A(\omega^{\prime}; \mathbf{k})\right]}{\left(i\omega_l+\mu-\omega\right)\left(i\omega_l-\Omega-i0+\mu-\omega^{\prime}\right)}\right),\\
\label{Kubo-L-21-A}
K_{nm}^{21} &=& -\frac{T}{e}\,\mbox{Re}\left(\lim_{\Omega\to0}\frac{i}{\Omega} T \sum_{l=-\infty}^{\infty} \int\frac{d^3\mathbf{k}}{(2\pi)^3}
\int \int d\omega d\omega^{\prime} \frac{\mbox{tr}\left[j_{n}^Q(\mathbf{k})A(\omega; \mathbf{k})j_{m}(\mathbf{k})
A(\omega^{\prime}; \mathbf{k})\right]}{\left(i\omega_l+\mu-\omega\right)\left(i\omega_l-\Omega-i0+\mu-\omega^{\prime}\right)}\right),\\
\label{Kubo-L-22-A}
K_{nm}^{22} &=& -T\,\mbox{Re}\left(\lim_{\Omega\to0}\frac{i}{\Omega} T \sum_{l=-\infty}^{\infty} \int\frac{d^3\mathbf{k}}{(2\pi)^3}
\int \int d\omega d\omega^{\prime} \frac{\mbox{tr}\left[j_{n}^Q(\mathbf{k})A(\omega; \mathbf{k})j_{m}^Q(\mathbf{k})
A(\omega^{\prime}; \mathbf{k})\right]}{\left(i\omega_l+\mu-\omega\right)\left(i\omega_l-\Omega-i0+\mu-\omega^{\prime}\right)}\right),
\end{eqnarray}
where $\omega_{l}=(2l+1)\pi T$ (with $l\in\mathbb{Z}$) are the fermionic Matsubara frequencies. In the derivation,
we used the spectral representation for the unperturbed Green's function
\begin{equation}
G^{(0)}(i\omega_l; \mathbf{k})=\int_{-\infty}^{\infty} d\omega \frac{A(\omega; \mathbf{k})}{i\omega_l+\mu-\omega},
\label{Kubo-Green-spectral-def}
\end{equation}
where the spectral function $A(\omega; \mathbf{k})$ is defined as usual in terms of the retarded and
advanced Green's functions,
\begin{equation}
A(\omega; \mathbf{k}) \equiv \frac{i}{2\pi}
\left[G^{(0)}(\omega+i0; \mathbf{k})-G^{(0)}(\omega-i0; \mathbf{k})\right]_{\mu=0} =
i\sum_{s=\pm}\frac{|\mathbf{d}|+s(\mathbf{d}\cdot\bm{\sigma})}{2|\mathbf{d}|} \delta\left(\omega-s|\mathbf{d}|\right).
\label{Kubo-spectral-function-def}
\end{equation}
As indicated by the $\delta$ function on the right-hand side, the spectral function $A(\omega; \mathbf{k})$ describes
noninteracting quasiparticles with the vanishing decay width. In realistic models, of course, the quasiparticle
decay width is generically nonzero. This can be implemented phenomenologically by replacing the delta-function
with a Lorentzian distribution, i.e.,
\begin{equation}
\delta_{\Gamma}(\omega-s|\mathbf{d}|)\equiv \frac{1}{\pi} \frac{\Gamma(\omega)}{(\omega-s|\mathbf{d}|)^2
+\Gamma^2(\omega)}.
\label{Kubo-d-Gamma}
\end{equation}
In this study we will use the following energy-dependent ansatz for the quasiparticle width $\Gamma(\omega)
=\Gamma_0(1 +\omega^2/\epsilon_0^2)$, where $\epsilon_0\equiv |\mathbf{d}|_{\mathbf{k}=\mathbf{0}}$ is a
characteristic energy scale set by the size of the ``dome" between the Weyl nodes, see Fig.~\ref{fig:model-energy}.
The ansatz for $\Gamma(\omega)$ is motivated, in part, by the study of Weyl semimetals with
a short-range disorder in Ref.~\cite{Burkov:2011}, which revealed a quadratic dependence of the quasiparticle
width on the energy, $\Gamma(\omega)\propto \omega^2$. In addition, we also included a nonzero constant
term $\Gamma_0$ in our model expression for $\Gamma(\omega)$. Such an extra term may mimic effects of other
types of disorder. For simplicity of the presentation, in the following we will omit the argument of $\Gamma$.

By making use of the formulas in Appendix~\ref{Sec:App-Matsubara}, we can easily perform the summations
over the Matsubara frequencies in Eqs.~(\ref{Kubo-L-11-A}) -- (\ref{Kubo-L-22-A}). Then, we will arrive at
the following expressions for the transport coefficients:
\begin{eqnarray}
\label{Kubo-L-11-A-1-alt}
K_{nm}^{11} &=& -\frac{1}{e^2}\,\mbox{Re}\Bigg(\lim_{\Omega\to0}\frac{i}{\Omega} \int\frac{d^3\mathbf{k}}{(2\pi)^3}
\int \int d\omega d\omega^{\prime} \frac{n_F(\omega)-n_F(\omega^{\prime})}{\omega-\omega^{\prime}-\Omega-i0} \mbox{tr}
\left[j_{n}(\mathbf{k})A(\omega; \mathbf{k})j_{m}(\mathbf{k})
A(\omega^{\prime}; \mathbf{k})\right]\Bigg),\\
\label{Kubo-L-12-A-1-alt}
K_{nm}^{12} &=& -\frac{T}{e}\,\mbox{Re}\Bigg(\lim_{\Omega\to0}\frac{i}{\Omega} \int\frac{d^3\mathbf{k}}{(2\pi)^3}
\int \int d\omega d\omega^{\prime} \frac{(\omega-\mu-\Omega)n_F(\omega)-(\omega^{\prime}-\mu)n_F(\omega^{\prime})}{\omega-\omega^{\prime}
-\Omega-i0} \mbox{tr}\left[j_{n}(\mathbf{k})A(\omega; \mathbf{k})j_{m}(\mathbf{k})
A(\omega^{\prime}; \mathbf{k})\right]\Bigg),\nonumber\\
\\
\label{Kubo-L-21-A-1-alt}
K_{nm}^{21} &=& -\frac{T}{e}\,\mbox{Re}\Bigg(\lim_{\Omega\to0}\frac{i}{\Omega}  \int\frac{d^3\mathbf{k}}{(2\pi)^3}
\int \int d\omega d\omega^{\prime} \frac{(\omega-\mu)n_F(\omega)-(\omega^{\prime}-\mu+\Omega)n_F(\omega^{\prime})}{\omega-\omega^{\prime}
-\Omega-i0} \mbox{tr}\left[ j_{n}(\mathbf{k})A(\omega; \mathbf{k})j_{m}(\mathbf{k})
A(\omega^{\prime}; \mathbf{k})\right]\Bigg),\nonumber\\
\\
\label{Kubo-L-22-A-1-alt}
K_{nm}^{22} &=& -T\,\mbox{Re}\Bigg(\lim_{\Omega\to0}\frac{i}{\Omega}\int\frac{d^3\mathbf{k}}{(2\pi)^3}
\int \int d\omega d\omega^{\prime} \frac{(\omega-\mu)(\omega-\mu-\Omega)n_F(\omega)-(\omega^{\prime}-\mu+\Omega)(\omega^{\prime}
-\mu)n_F(\omega^{\prime})}{\omega-\omega^{\prime}-\Omega-i0} \nonumber\\
&\times&\mbox{tr}\left[j_{n}(\mathbf{k})A(\omega; \mathbf{k}) j_{m}(\mathbf{k})
A(\omega^{\prime}; \mathbf{k})\right]\Bigg),
\end{eqnarray}
where $n_{F}(\omega)=1/\left[e^{(\omega-\mu)/T}+1\right]$ is the Fermi--Dirac distribution function.

\subsection{Electromagnetic orbital and heat magnetizations}
\label{sec:Kubo-magnetization-all}

The electromagnetic orbital magnetization $\mathbf{M}$ can be calculating by inverting the Streda formula \cite{Streda:1982},
\begin{equation}
\label{Kubo-magnetization-Streda}
\sigma_{nm}^{II} = -e\epsilon_{nml} \frac{\partial M_l}{\partial\mu},
\end{equation}
where $\sigma_{nm}^{II}$ denotes the thermodynamical part of the electric conductivity, originating from
filled states below the Fermi level.

By making use of the Kubo--Streda formalism \cite{Streda:1982}, one can derive the following formal result for
the electric conductivity tensor (see, e.g., Refs.~\cite{Yang-Chang:2006,Nunner:2007}):
\begin{eqnarray}
\label{Kubo-magnetization-sigma-II}
\sigma_{nm}^{II} &=& -\frac{1}{4\pi} \mbox{Re}\Bigg(\int \frac{d^3 \mathbf{k}}{(2\pi)^3} \int_{-\infty}^{\infty}
d \omega\,n_{F}(\omega)\, \mbox{tr}\Bigg[j_n(\mathbf{k})G^{(0)}(\omega-\mu+i0;\mathbf{k})j_m(\mathbf{k})
\left(\partial_{\omega}G^{(0)}(\omega-\mu+i0;\mathbf{k})\right) \nonumber\\
&-&j_n(\mathbf{k})\left(\partial_{\omega}G^{(0)}(\omega-\mu+i0;\mathbf{k})\right)j_m(\mathbf{k})
G^{(0)}(\omega-\mu+i0;\mathbf{k}) - H.c.\Bigg]\Bigg).
\end{eqnarray}
Here all diagonal components of the above tensor vanish. Now, by using the explicit expression for
the Green's function in the clean limit given by Eq.~(\ref{model-G0-def-be}) and calculating the trace, we obtain
\begin{eqnarray}
\label{Kubo-magnetization-sigma-II-2}
\sigma_{nm}^{II} &=& -\frac{e^2}{\pi} \mbox{Re}\Bigg(\int \frac{d^3 \mathbf{k}}{(2\pi)^3} \int_{-\infty}^{\infty}
d \omega\,n_{F}(\omega)\, i\, \sum_{s=\pm} \Omega_{nm}(\mathbf{k})
\left[\frac{-1}{\left(\omega-s|\mathbf{d}|+i0\right)\left(\omega+s|\mathbf{d}|+i0\right)}
+\frac{1}{\left(\omega-s|\mathbf{d}| +i0\right)^2}\right] \Bigg) \nonumber\\
&=& e^2 \int \frac{d^3 \mathbf{k}}{(2\pi)^3}
\Omega_{nm}(\mathbf{k})
\left\{ \left[n_F(|\mathbf{d}|)-n_F(-|\mathbf{d}|)\right] - |\mathbf{d}|\left[n_F^{\prime}(|\mathbf{d}|)
+n_F^{\prime}(-|\mathbf{d}|)\right]\right\},
\end{eqnarray}
where we integrated by parts to obtain the second term in the curly brackets and introduced the
following Berry curvature tensor:
\begin{equation}
\Omega_{nm}(\mathbf{k})=\frac{1}{2|\mathbf{d}|^3}
\left(\mathbf{d}\cdot \left[(\partial_{k_n}\mathbf{d})\times (\partial_{k_m}\mathbf{d})\right] \right).
\label{Kubo-Berry-tensor}
\end{equation}
Since the magnetization should vanish in the limit $\mu\to -\infty$ \cite{Qin-Niu:2011}, we can integrate
the relation in Eq.~(\ref{Kubo-magnetization-Streda}) and obtain the following result:
\begin{equation}
\label{Kubo-magnetization-Streda-int-2}
M_l = -\frac{\epsilon_{nml}}{2e}\int_{-\infty}^{\mu}d\mu_0\sigma_{nm}^{II}(\mu_0)
=-e \,\frac{\epsilon_{nml}}{2} \int \frac{d^3 \mathbf{k}}{(2\pi)^3} \Omega_{nm}(\mathbf{k})
\left\{ T\ln{\left(\frac{1+e^{(\mu-|\mathbf{d}|)/T}}{1+e^{(\mu+|\mathbf{d}|)/T}}\right)}
+ |\mathbf{d}|\left[n_F(|\mathbf{d}|)+n_F(-|\mathbf{d}|)\right] \right\}.
\end{equation}
By noting that the expression on the right-hand side contains the Berry curvature tensor,
we conclude that this magnetization has a topological origin. This becomes even more
transparent in the limit of small chemical potential and zero temperature, i.e.,
\begin{equation}
\label{Kubo-magnetization-Streda-int-00}
M_l \simeq e \mu \frac{\epsilon_{nml}}{2} \int \frac{d^3 \mathbf{k}}{(2\pi)^3} \Omega_{nm}(\mathbf{k})
= n \frac{e \mu b_l}{2\pi^2},
\end{equation}
where the result is determined by the same winding number of the mapping of a two dimensional section
of the Brillouin zone onto a unit sphere as the electric Bardeen--Zumino current in Ref.~\cite{Gorbar:2017-Bardeen}.

The heat magnetization $\mathbf{M}^Q$ can be calculated in a similar way. The starting point is the
Streda-like formula for the heat magnetization:
\begin{equation}
\label{Kubo-magnetization-heat-Streda}
\sigma_{nm}^{II,Q} = -e\epsilon_{nml} \frac{\partial M_l^Q}{\partial\mu},
\end{equation}
where the tensor $\sigma_{nm}^{II,Q}$ is defined by the mixed current-current correlator,
\begin{eqnarray}
\label{Kubo-magnetization-heat-sigma-II}
\sigma_{nm}^{II,Q} &=& -\frac{1}{4\pi} \mbox{Re}\Bigg(\int \frac{d^3 \mathbf{k}}{(2\pi)^3} \int_{-\infty}^{\infty} d \omega\,n_{F}(\omega)\,
\mbox{tr}\Bigg[j_n(\mathbf{k})G^{(0)}(\omega-\mu+i0;\mathbf{k})j_m^Q(\omega-\mu;\mathbf{k}) \left(\partial_{\omega}G^{(0)}(\omega-\mu+i0;\mathbf{k})\right)
\nonumber\\
&-&j_n(\mathbf{k})\left(\partial_{\omega}G^{(0)}(\omega-\mu+i0;\mathbf{k})\right)j_m^Q(\omega-\mu;\mathbf{k}) G^{(0)}(\omega-\mu+i0;\mathbf{k})
- H.c.\Bigg]\Bigg).
\end{eqnarray}
By making use of the explicit expression for the Green's function (\ref{model-G0-def-be}) and integrating over the
energy $\omega$, we derive
\begin{eqnarray}
\label{Kubo-magnetization-heat-sigma-II-2}
\sigma_{nm}^{II,Q} &=& e \int \frac{d^3 \mathbf{k}}{(2\pi)^3} \Omega_{nm}(\mathbf{k})
\Bigg\{ \left[(|\mathbf{d}|-\mu)n_F(|\mathbf{d}|)+(|\mathbf{d}|+\mu)n_F(-|\mathbf{d}|)\right]
\nonumber\\
&-&
|\mathbf{d}|
\left[\partial_{\omega}(\omega-\mu)n_F(\omega)\right]\Big|_{\omega\to|\mathbf{d}|} - |\mathbf{d}|
\left[\partial_{\omega}(\omega-\mu)n_F(\omega)\right]\Big|_{\omega\to-|\mathbf{d}|}\Bigg\}.
\end{eqnarray}
Following the same approach as in the derivation of the electromagnetic orbital magnetization, we
integrate the relation in Eq.~(\ref{Kubo-magnetization-heat-Streda}) over $\mu$ and arrive at the
final result for the heat magnetization
\begin{eqnarray}
\label{Kubo-magnetization-heat-Streda-int-2}
M_l^Q &=& -\frac{\epsilon_{nml}}{2e}\int_{-\infty}^{\mu}d\mu_0\sigma_{nm}^{II,Q}(\mu_0)
=\frac{\epsilon_{nml}}{2} \int \frac{d^3 \mathbf{k}}{(2\pi)^3} \Omega_{nm}(\mathbf{k})
\Bigg\{
T(\mu-|\mathbf{d}|)\ln{\left(1+e^{(\mu-|\mathbf{d}|)/T}\right)} \nonumber\\
&-&T(\mu+|\mathbf{d}|)
\ln{\left(1+e^{(\mu+|\mathbf{d}|)/T}\right)} +T^2\mbox{Li}_2\left(-e^{(\mu-|\mathbf{d}|)/T}\right)
-T^2\mbox{Li}_2\left(-e^{(\mu+|\mathbf{d}|)/T}\right) \nonumber\\
&-&|\mathbf{d}|
\left[(|\mathbf{d}|-\mu)n_F(|\mathbf{d}|)-(|\mathbf{d}|+\mu)n_F(-|\mathbf{d}|)\right]
\Bigg\}.
\end{eqnarray}
%The results of this section for the Kubo's transport coefficients and the magnetizations are used in the next sections to analyze the thermoelectric transport of multi-Weyl semimetals.

Before concluding this section, let us mention that the results in
Eqs.~(\ref{Kubo-magnetization-Streda-int-2}) and (\ref{Kubo-magnetization-heat-Streda-int-2})
have a topological origin. This is evident from the fact that the corresponding expressions
contain the Berry curvature in their integrands. As we will see below, the tensor structure of
these magnetizations is the same as that of the nondissipative parts of the Kubo's coefficients.
This is not accidental, however, since the latter have a similar topological origin.

\section{Thermoelectric transport coefficients $L_{nm}^{\alpha \beta}$}
\label{sec:Kubo-L-All}

By using the results for the Kubo's transport coefficients and the magnetizations from the previous section, here we obtain the Kubo's response coefficients $K_{nm}^{\alpha \beta}$ and then
provide the results for the thermoelectric transport coefficients $L_{nm}^{\alpha \beta}$.

\subsection{Coefficient $L_{nm}^{11}$}
\label{sec:Kubo-L11}

The transport coefficient $L_{nm}^{11}=K_{nm}^{11}\equiv \sigma_{nm}/e^2$ describes the electric
conductivity. The corresponding conductivity tensor $\sigma_{nm}$ was calculated by us in the same
lattice model in Ref.~\cite{Gorbar:2017-Bardeen}. Therefore, here we provide only the final result generalized to the case of nonzero temperature, i.e.,
\begin{equation}
\label{Kubo-L-11-A-2}
L_{nm}^{11} =L_{nm}^{11, {\rm D}} +L_{nm}^{11, {\rm ND}},
\end{equation}
where the dissipative and nondissipative parts of the corresponding transport coefficient are given by
\begin{eqnarray}
\label{Kubo-L-11-A-2-R}
L_{nm}^{11, {\rm D}} &=& 2 \pi\, \int\frac{d^3\mathbf{k}}{(2\pi)^3} \int  \frac{d\omega }{4T\cosh^2{\left(\frac{\omega-\mu}{2T}\right)}}
\sum_{s,s^{\prime}=\pm}\frac{ss^{\prime}}{4|\mathbf{d}|^2}\delta_{\Gamma}\left(\omega-s|\mathbf{d}|\right)\delta_{\Gamma}
\left(\omega-s^{\prime}|\mathbf{d}|\right) \nonumber\\
&\times&
\Big[(ss^{\prime}-1) |\mathbf{d}|^2\left(\left(\partial_{k_n}\mathbf{d})\cdot
(\partial_{k_m}\mathbf{d}\right)\right)
+2\left(\mathbf{d}\cdot(\partial_{k_n}\mathbf{d})\right)\left(\mathbf{d}\cdot(\partial_{k_m}\mathbf{d})\right) \Big],
\end{eqnarray}
and
\begin{eqnarray}
\label{Kubo-L-11-A-2-I-gamma-not-0}
L_{nm}^{11, {\rm ND}} =
4\int\frac{d^3\mathbf{k}}{(2\pi)^3} \int \int d\omega d\omega^{\prime} \frac{\left[n_F(\omega)-n_F(\omega^{\prime})\right]}
{(\omega-\omega^{\prime})^2}|\mathbf{d}|^2
\delta_{\Gamma}\left(\omega-|\mathbf{d}|\right)\delta_{\Gamma}\left(\omega^{\prime}+|\mathbf{d}|\right) \Omega_{nm}(\mathbf{k}),
\end{eqnarray}
respectively. As is easy to check, the only nonzero components of the dissipative part are $L_{xx}^{11, {\rm D}}
=L_{yy}^{11, {\rm D}}$ and $L_{zz}^{11, {\rm D}}$. They describe the electric charge transport in the transverse
and longitudinal directions with respect to the chiral shift $\mathbf{b}$. The corresponding components of the
conductivity tensor are $\sigma_{xx}=\sigma_{yy}\equiv e^2 L_{xx}^{11, {\rm D}}$ and $\sigma_{zz}\equiv e^2
L_{zz}^{11, {\rm D}}$.

By noting that the integrand on the right-hand side of Eq.~(\ref{Kubo-L-11-A-2-I-gamma-not-0}) is proportional
to the Berry curvature, we conclude that the nondissipative part has a topological origin. In the lattice model
used, the only nontrivial components of the corresponding antisymmetric tensor are $L_{xy}^{11, {\rm ND}}
=-L_{yx}^{11, {\rm ND}}$. They remain finite even in the clean limit $\Gamma\to0$ and describe the anomalous
Hall effect. Therefore, for simplicity, in the following we will consider these nondissipative terms only in the
clean limit, i.e.,
\begin{equation}
\label{Kubo-L-11-A-2-I}
\lim_{\Gamma\to0}L_{nm}^{11, {\rm ND}}
= \int\frac{d^3\mathbf{k}}{(2\pi)^3} \left[n_F(|\mathbf{d}|)-n_F(-|\mathbf{d}|)\right] \Omega_{nm}(\mathbf{k}).
\end{equation}
As is easy to check, in the limit of zero temperature $T\to0$ and vanishing chemical potential $\mu=0$,
this leads to the well-known result for the anomalous Hall conductivity
\cite{Ran,Burkov:2011ene,Grushin-AHE,Zyuzin,Goswami,1705.04576}:
\begin{equation}
\sigma_{\text{{\tiny AHE}}}\equiv  \lim_{T\to 0}\lim_{\mu\to 0} e^2 L_{xy}^{11}
= -n\frac{e^2b_z}{2\pi^2}.
\label{Kubo-E-AHE}
\end{equation}
In terms of the currents, this corresponds to the topological Bardeen--Zumino contribution
$J_{\text{{\tiny BZ}}} =  -n e^2[\mathbf{E}\times\mathbf{b}]/(2\pi^2)$ \cite{Gorbar:2017-Bardeen}
(see also Refs.~\cite{Landsteiner:2013sja,Landsteiner:2016} for the related discussions in the
case of $n=1$ Weyl semimetals).

\begin{figure}[t]
\begin{center}
\hspace{-0.32\textwidth}(a)\hspace{0.32\textwidth}(b)\hspace{0.32\textwidth}(c)\\[0pt]
\includegraphics[width=0.32\textwidth]{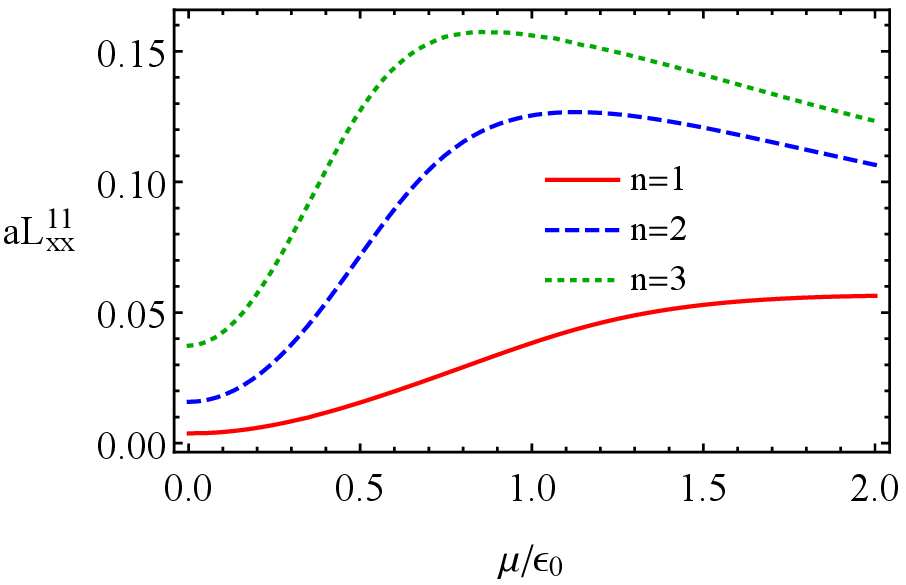}\hfill
\includegraphics[width=0.32\textwidth]{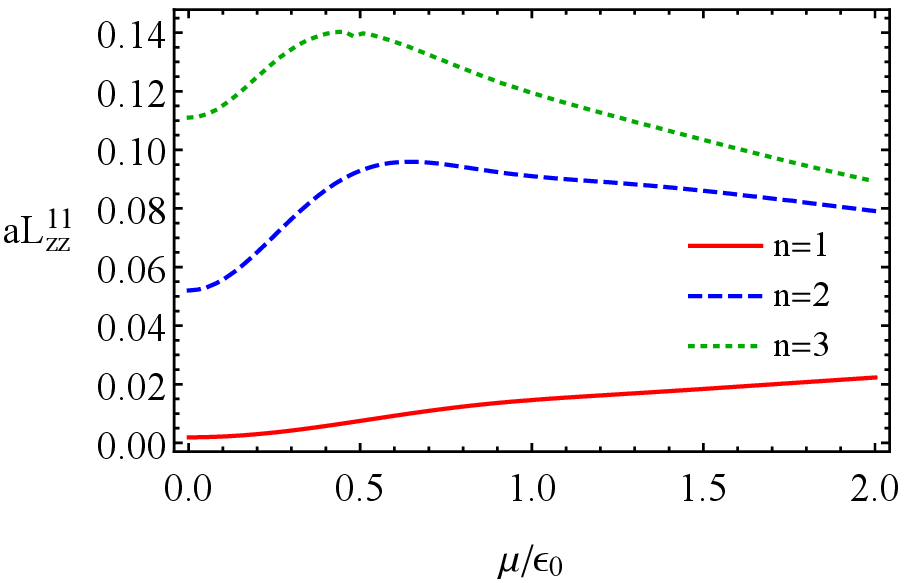}\hfill
\includegraphics[width=0.32\textwidth]{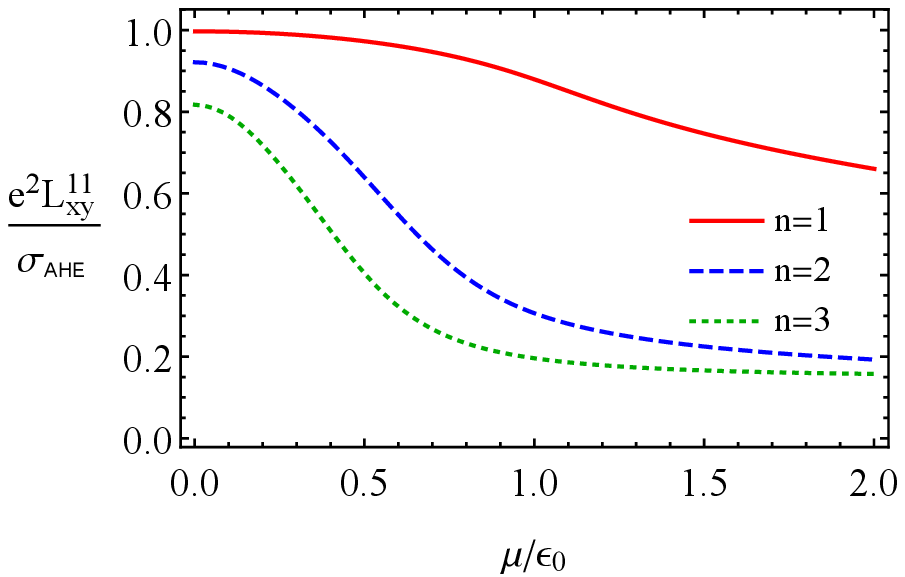}
\caption{The dependence of the transport coefficients $L_{xx}^{11}$, $L_{zz}^{11}$, and $L_{xy}^{11}$ on
the chemical potential in a Weyl semimetal (red solid line), a double-Weyl semimetal (blue dashed line),
and a triple-Weyl semimetal (green dotted line) at fixed $T=0.1\,\epsilon_0$. In panels (a) and (b),
the quasiparticle transport width is modeled by $\Gamma=\Gamma_0\left(1+\omega^2/\epsilon_0^2\right)$
with $\Gamma_0=0.1\,\epsilon_0$. In panel (c), the results are plotted in the clean limit, $\Gamma=0$.
The numerical values of other model parameters are defined in Appendix~\ref{Sec:App-model}.}
\label{fig:Kubo-E-L11-mu}
\end{center}
\end{figure}

For multi-Weyl semimetals with $n=1,2,3$, the dependence of the transport coefficients
$L_{xx}^{11}$, $L_{zz}^{11}$, and $L_{xy}^{11}$ on the chemical potential is shown in
Fig.~\ref{fig:Kubo-E-L11-mu}. The corresponding numerical results are obtained at a small, but nonzero
temperature, $T=0.1\,\epsilon_0$. We used the quasiparticle transport width $\Gamma=\Gamma_0
\left(1+\omega^2/\epsilon_0^2\right)$ with $\Gamma_0=0.1\,\epsilon_0$ in the calculation of the dissipative transport coefficients,
shown in Figs.~\ref{fig:Kubo-E-L11-mu}(a) and \ref{fig:Kubo-E-L11-mu}(b), and set $\Gamma=0$ in the calculation of the nondissipative
transport coefficient, shown in Fig.~\ref{fig:Kubo-E-L11-mu}(c). The numerical values of other parameters of our
model are defined in Appendix~\ref{Sec:App-model}.

%%%%%%%%%%%%%%%%%%
%%%%%%%%%%%%%%%%%%

As we see from Fig.~\ref{fig:Kubo-E-L11-mu}, all multi-Weyl semimetals with $n=1,2,3$ share a
similar behavior of their transport coefficients $L_{nm}^{11}$ as functions of $\mu$. We note, however,
that the Weyl materials with larger values of $n$ tend to have a steeper dependence on the chemical
potential in the region of small $\mu$. We also find that the dissipative coefficients $L_{xx}^{11}$ and
$L_{zz}^{11}$ tend to be more nonmonotonous in the double- and triple-Weyl semimetals than in the
$n=1$ Weyl semimetals. While this feature appears to be quite robust in the model
used, it is hard to say how generic it is in reality. By noting that the maximum values of $L_{xx}^{11}$
and $L_{zz}^{11}$ are obtained at $\mu \sim \epsilon_0$, one might suggest that the nonmonotonic
behavior is connected with qualitative changes in the density of states near/above the Lifshitz transition
in the present model. In realistic materials, however, the band structures are much more complicated than
in our model and, therefore, the above predictions are hard to justify away from the region of small
$\mu$.

\begin{figure}[t]
\begin{center}
\hspace{-0.32\textwidth}(a)\hspace{0.32\textwidth}(b)\hspace{0.32\textwidth}(c)\\[0pt]
\includegraphics[width=0.32\textwidth]{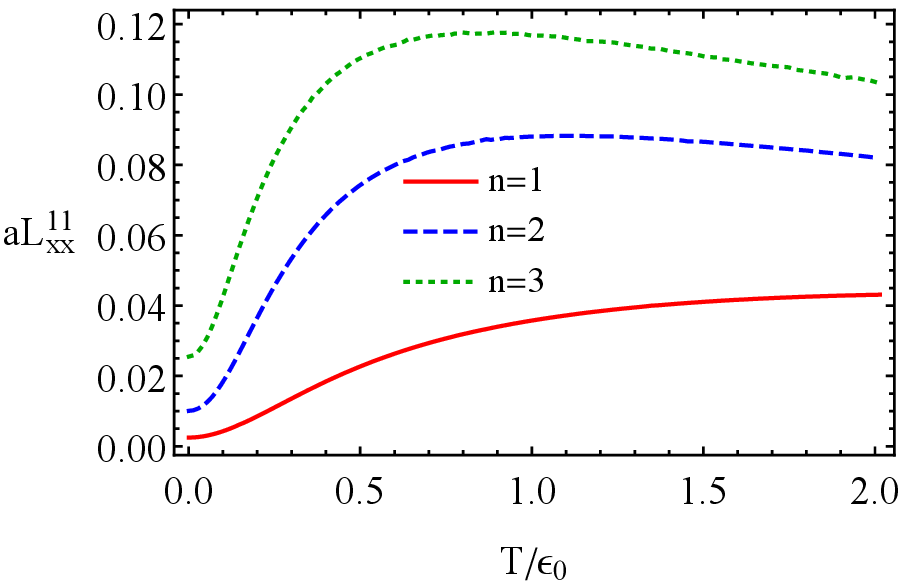}\hfill
\includegraphics[width=0.32\textwidth]{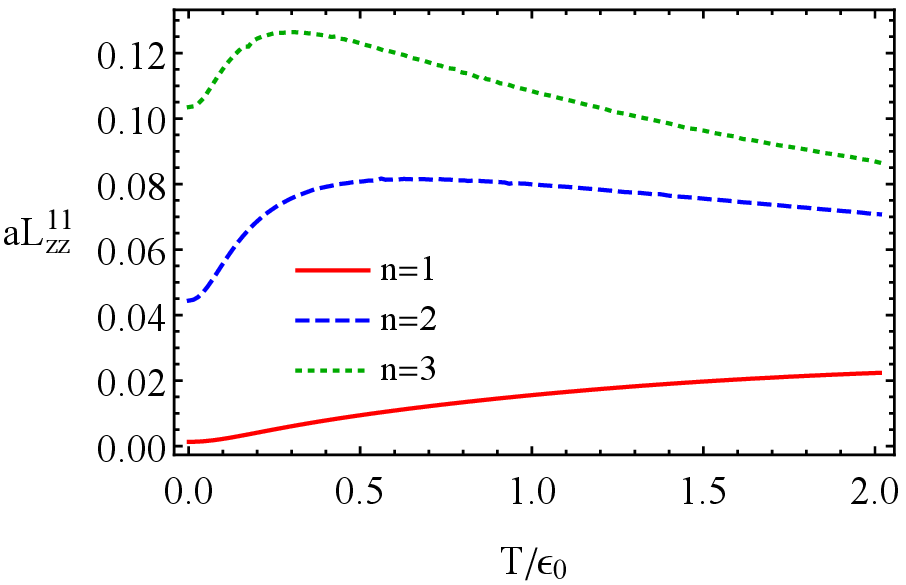}\hfill
\includegraphics[width=0.32\textwidth]{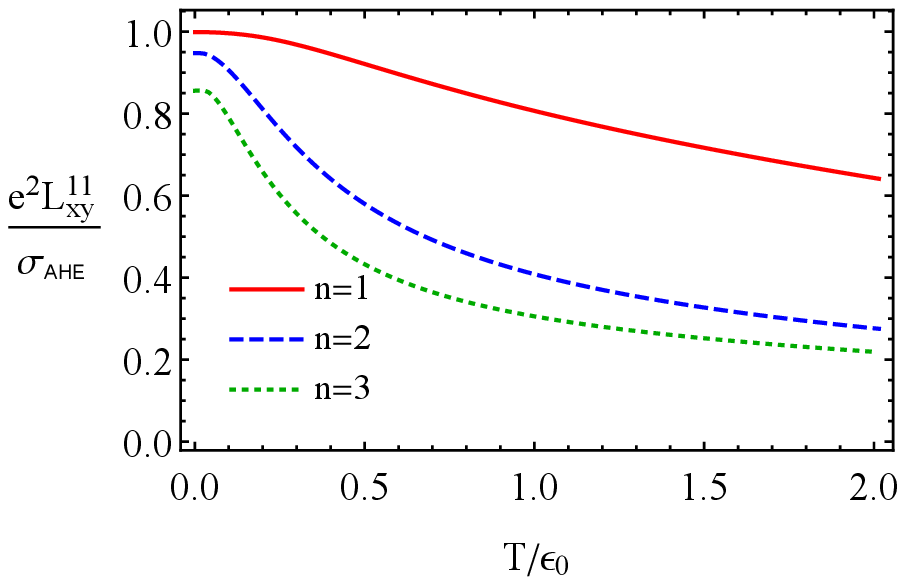}
\caption{The dependence of the transport coefficients $L_{xx}^{11}$, $L_{zz}^{11}$, and $L_{xy}^{11}$ on the
temperature in a Weyl semimetal (red solid line), a double-Weyl semimetal (blue dashed line), and a triple-Weyl
semimetal (green dotted line) at fixed $\mu=0.1\,\epsilon_0$. In panels (a) and (b), the quasiparticle
transport width is modeled by $\Gamma=\Gamma_0\left(1+\omega^2/\epsilon_0^2\right)$ with
$\Gamma_0=0.1\,\epsilon_0$. In panel (c), the results are plotted in the clean limit, $\Gamma=0$.
The numerical values of other model parameters are defined in Appendix~\ref{Sec:App-model}.}
\label{fig:Kubo-E-L11-T}
\end{center}
\end{figure}

As we see Fig.~\ref{fig:Kubo-E-L11-mu}(c), the results for the anomalous Hall
conductivity are slightly smaller than $\sigma_{\text{{\tiny AHE}}}$ in Eq.~(\ref{Kubo-E-AHE}) even when the
chemical potential approaches zero. This is due to the fact that we fixed a small, but nonzero temperature
$T=0.1\,\epsilon_0$ when presenting the results. In this connection, we note that the anomalous Hall
conductivity generically decreases with increasing $\mu$ and/or $T$ [see also
Fig.~\ref{fig:Kubo-E-L11-T}(c)]. The corresponding dependence is again much steeper in multi-Weyl
semimetals with higher $n$.

%%%%%%%%%%%%%%%%%%
%%%%%%%%%%%%%%%%%%

The temperature dependence of the same three transport coefficients is shown in Fig.~\ref{fig:Kubo-E-L11-T}
for a fixed value of the chemical potential $\mu=0.1\,\epsilon_0$. As expected, the dissipative coefficients
$L_{xx}^{11}$ and $L_{zz}^{11}$ for multi-Weyl semimetals with $n=1,2,3$ are nonmonotonic
functions of temperature. We see that the slopes generically increase with the value of the topological
charge $n$. In connection to the anomalous Hall conductivity, shown in 
Fig.~\ref{fig:Kubo-E-L11-T}(c), we note that the results differ slightly from $\sigma_{\text{{\tiny AHE}}}$ in
Eq.~(\ref{Kubo-E-AHE}) even in the limit $T\to0$. The deviation comes from the fact that a nonzero
value of the chemical potential $\mu=0.1\,\epsilon_0$ was used in the calculation. As expected,
increasing the temperature tends to gradually wash away the anomalous Hall effect.

Before concluding the discussion of the electric conductivity, let us
compare the diagonal components of $\sigma_{nm}$ from the
Kubo's formalism with those obtained in the linearized chiral kinetic (Boltzmann)
theory \cite{Lundgren:2014hra}. At low temperatures, the latter predicts the
following behavior:
\begin{equation}
\sigma_{xx}\propto \frac{1}{\Gamma} \left(\mu^2 + \sigma_0 T^2\right),
\label{Kubo-L-11-sigma-CKT}
\end{equation}
where $\sigma_0$ is a numerical coefficient. (By assuming that the temperature is sufficiently
low, one can replace $\omega$ with $\mu$ in the expression for $\Gamma$.)
As one can check, such a dependence on $\mu$ and $T$ agrees well with our results
in Fig.~\ref{fig:Kubo-E-L11-mu}(a), as well as Fig.~\ref{fig:Kubo-E-L11-T}(a)
at sufficiently low temperatures.

\subsection{Coefficient $L_{nm}^{21}$}
\label{sec:Kubo-L21}

As is clear from Eq.~(\ref{Kubo-Je-Jq-gen-ee}), the flow of the heat current in response to the
external electric field $\mathbf{E}$ is quantified by the transport coefficient
$L_{nm}^{21}=K_{nm}^{21}-T\epsilon_{nml}M_l/e$, where the associated Kubo's coefficient
is defined by Eq.~(\ref{Kubo-L-21-A-1-alt}) and the orbital magnetization $\mathbf{M}$ is
given by Eq.~(\ref{Kubo-magnetization-Streda-int-2}).

After calculating the trace in Eq.~(\ref{Kubo-L-21-A-1-alt}), we find that the expression
for the Kubo's coefficient contains dissipative and nondissipative parts, i.e.,
\begin{eqnarray}
\label{Kubo-L-21-A-2-alt}
K_{nm}^{21} =K_{nm}^{21, {\rm D}} +K_{nm}^{21, {\rm ND}},
\end{eqnarray}
where
\begin{eqnarray}
\label{Kubo-L-21-A-2-R-alt}
K_{nm}^{21, {\rm D}}
&=& 2\pi\,T \int\frac{d^3\mathbf{k}}{(2\pi)^3} \int d\omega
\frac{\omega-\mu}{4T\cosh^2{\left(\frac{\omega-\mu}{2T}\right)}} \sum_{s,s^{\prime}
=\pm}\frac{ss^{\prime}}{4|\mathbf{d}|^2}\delta_{\Gamma}\left(\omega-s|\mathbf{d}|\right)\delta_{\Gamma}
\left(\omega-s^{\prime}|\mathbf{d}|\right)\nonumber\\
&\times&
\Big[(ss^{\prime}-1) |\mathbf{d}|^2\left(\left(\partial_{k_n}\mathbf{d})\cdot(\partial_{k_m}\mathbf{d}\right)\right)
+2\left(\mathbf{d}\cdot(\partial_{k_n}\mathbf{d})\right)\left(\mathbf{d}\cdot(\partial_{k_m}\mathbf{d})\right) \Big]
\end{eqnarray}
and
\begin{eqnarray}
\label{Kubo-L-21-A-2-I-gamma-not-0}
K_{nm}^{21, {\rm ND}} &=&
T\int\frac{d^3\mathbf{k}}{(2\pi)^3} \int \int d\omega d\omega^{\prime} \frac{\left[n_F(\omega)-n_F(\omega^{\prime})\right]
(\omega+\omega^{\prime}-2\mu)}{(\omega-\omega^{\prime})^2} 2|\mathbf{d}|^2
\delta_{\Gamma}\left(\omega-|\mathbf{d}|\right)\delta_{\Gamma}\left(\omega^{\prime}+|\mathbf{d}|\right) \Omega_{nm}(\mathbf{k}).
\end{eqnarray}
It is straightforward to check that the only nontrivial components of the nondissipative part are
$K_{xy}^{21, {\rm ND}} = -K_{yx}^{21, {\rm ND}}$. These are topological terms that remain finite
even in the clean limit $\Gamma\to0$. Therefore, by following the same assumptions as in the
calculation of the magnetization, below we will consider these nondissipative terms in the clean
limit, i.e.,
\begin{eqnarray}
\label{Kubo-L-21-A-2-I-alt}
\lim_{\Gamma\to0}K_{nm}^{21, {\rm ND}}
= -T\, \mu \int\frac{d^3\mathbf{k}}{(2\pi)^3} \left[n_F(|\mathbf{d}|)-n_F(-|\mathbf{d}|)\right]
\Omega_{nm}(\mathbf{k}).
\end{eqnarray}
By combining the results for the Kubo's coefficients in Eqs.~(\ref{Kubo-L-21-A-2-R-alt}) and
(\ref{Kubo-L-21-A-2-I-alt}) with the magnetization in Eq.~(\ref{Kubo-magnetization-Streda-int-2}),
we can now calculate the thermoelectric transport coefficient $L_{nm}^{21}=K_{nm}^{21}
-T\epsilon_{nml}M_l/e$. As is easy to check, the only nonzero components of tensor $L_{nm}^{21}$
are $L_{xx}^{21}=L_{yy}^{21}$, $L_{zz}^{21}$, and $L_{xy}^{21}=-L_{yx}^{21}$.

\begin{figure}[t]
\begin{center}
\hspace{-0.32\textwidth}(a)\hspace{0.32\textwidth}(b)\hspace{0.32\textwidth}(c)\\[0pt]
\includegraphics[width=0.32\textwidth]{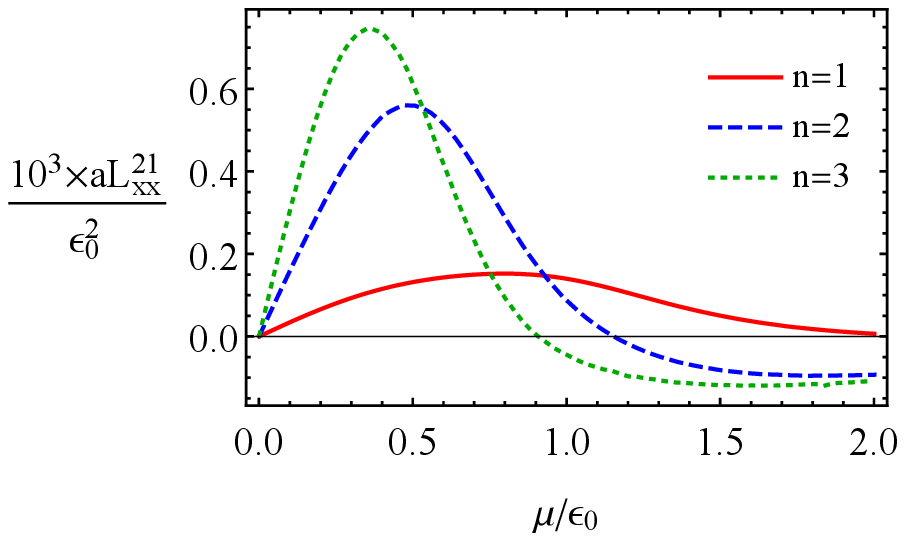}\hfill
\includegraphics[width=0.32\textwidth]{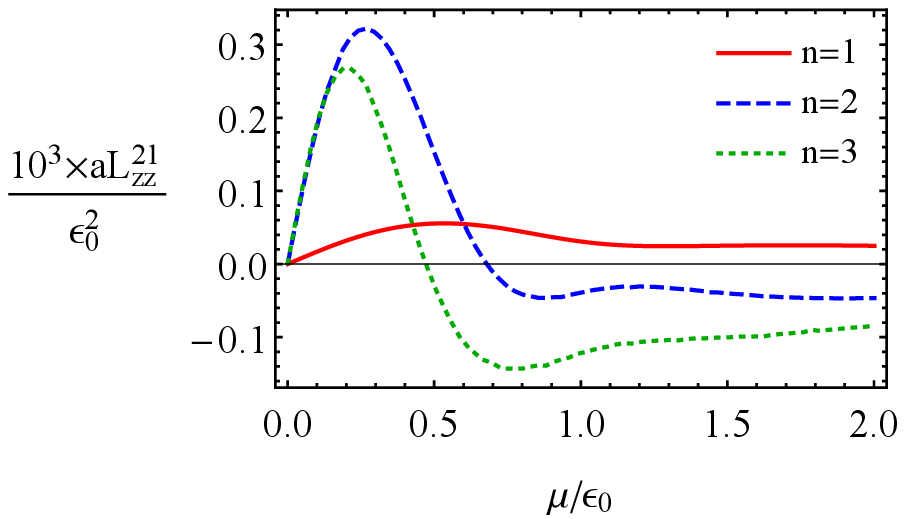}\hfill
\includegraphics[width=0.32\textwidth]{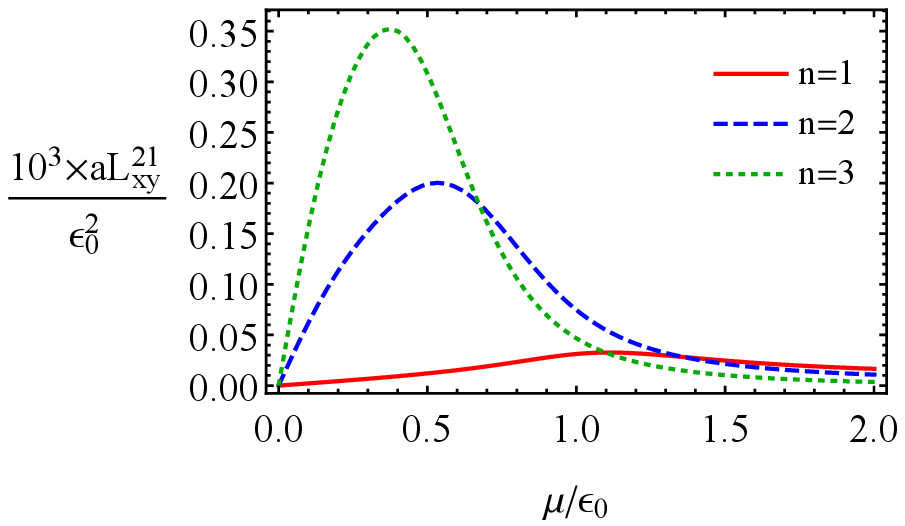}
\caption{The dependence of the transport coefficients $L_{xx}^{21}$, $L_{zz}^{21}$, and $L_{xy}^{21}$ on
the chemical potential in a Weyl semimetal (red solid line), a double-Weyl semimetal (blue dashed line), and
a triple-Weyl semimetal (green dotted line) at fixed $T=0.1\,\epsilon_0$. In panels (a) and (b), the
quasiparticle transport width is modeled by $\Gamma=\Gamma_0\left(1+\omega^2/\epsilon_0^2\right)$ with
$\Gamma_0=0.1\,\epsilon_0$. In panel (c), the results are plotted in the clean limit, $\Gamma=0$.
The numerical values of other model parameters are defined in Appendix~\ref{Sec:App-model}.}
\label{fig:Kubo-E-L21-mu}
\end{center}
\end{figure}

The dependence of the transport coefficients $L_{xx}^{21}$, $L_{zz}^{21}$, and $L_{xy}^{21}$ on the
chemical potential at fixed temperature $T=0.1\,\epsilon_0$ is presented in Fig.~\ref{fig:Kubo-E-L21-mu}
for multi-Weyl semimetals with different values of the topological charge $n=1,2,3$.
As in the rest of this paper, we plot the results for the
dissipative parts $L_{xx}^{21}$, $L_{zz}^{21}$ using the model of quasiparticles with nonzero width
$\Gamma=\Gamma_0\left(1+\omega^2/\epsilon_0^2\right)$. In contrast, the results for the nondissipative
coefficient $L_{xy}^{21}$ are presented in the clean limit, $\Gamma\to0$. Note that these are the same
assumptions that we used in the calculation of the electrical conductivity in the previous subsection.

\begin{figure}[t]
\begin{center}
\hspace{-0.32\textwidth}(a)\hspace{0.32\textwidth}(b)\hspace{0.32\textwidth}(c)\\[0pt]
\includegraphics[width=0.32\textwidth]{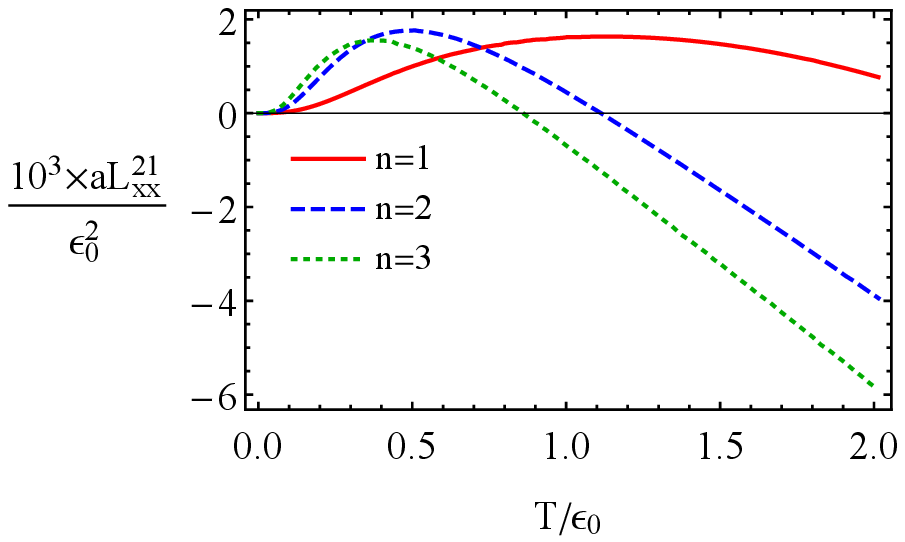}\hfill
\includegraphics[width=0.32\textwidth]{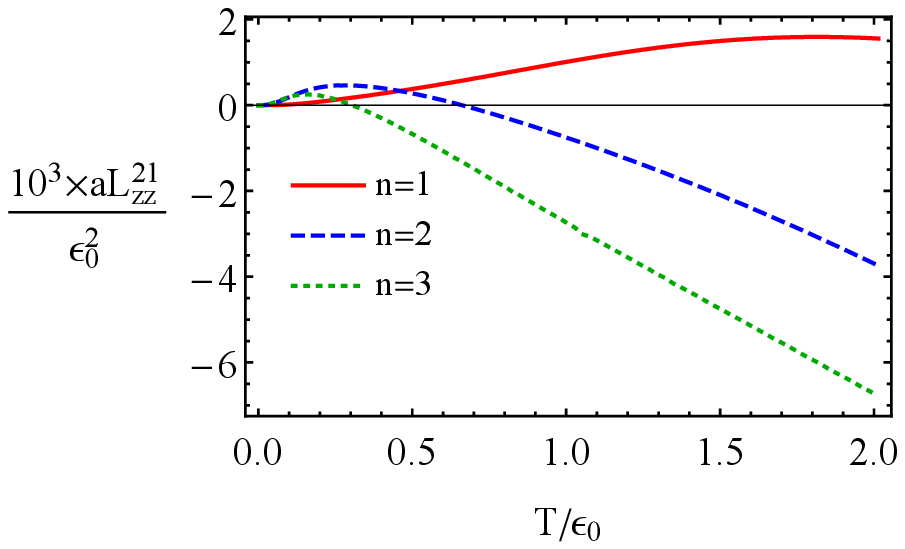}\hfill
\includegraphics[width=0.32\textwidth]{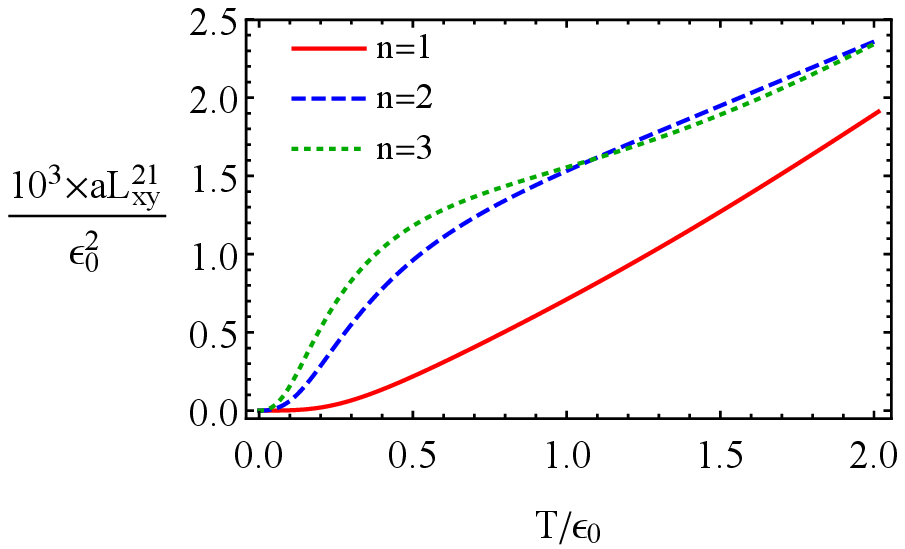}
\caption{The dependence of the transport coefficients $L_{xx}^{21}$, $L_{zz}^{21}$, and $L_{xy}^{21}$ on
temperature in a Weyl semimetal (red solid line), a double-Weyl semimetal (blue dashed line), and
a triple-Weyl semimetal (green dotted line) at fixed $\mu=0.1\,\epsilon_0$. In panels (a) and (b), the
quasiparticle transport width is modeled by $\Gamma=\Gamma_0\left(1+\omega^2/\epsilon_0^2\right)$ with
$\Gamma_0=0.1\,\epsilon_0$. In panel (c), the results are plotted in the clean limit, $\Gamma=0$.
The numerical values of other model parameters are defined in Appendix~\ref{Sec:App-model}.}
\label{fig:Kubo-E-L21-T}
\end{center}
\end{figure}

As the results in Fig.~\ref{fig:Kubo-E-L21-mu} demonstrate, all three transport coefficients are nonmonotonic
functions of $\mu$. Moreover, as we see from Figs.~\ref{fig:Kubo-E-L21-mu}(a) and \ref{fig:Kubo-E-L21-mu}(b),
the dissipative parts $L_{xx}^{21}$ (transverse electrothermal coefficient) and  $L_{zz}^{21}$ (longitudinal electrothermal
coefficient) in the double-Weyl (dashed blue lines) and triple-Weyl (dotted green lines) semimetals change
their signs at sufficiently large values of the chemical potential, $\mu\sim \epsilon_0$. This is in contrast
to the situation in Weyl semimetals with the topological charge $n=1$ (solid red lines), where the
corresponding coefficients remain positive at given values of $\mu$. Moreover, a similar qualitative behavior with the
change of sign at $T\sim \epsilon_0$ is also observed in the temperature dependence of these
coefficients. The corresponding results are shown Figs.~\ref{fig:Kubo-E-L21-T}(a) and \ref{fig:Kubo-E-L21-T}(b).

%%%%%%%%%%%%%%%%%%
%%%%%%%%%%%%%%%%%%

%%%%%%%%%%%%%%%%%%
%%%%%%%%%%%%%%%%%%

Guided by our findings, it might be tempting to suggest that the change of sign in the dissipative electrothermal
coefficients at sufficiently large chemical potentials and/or temperatures is a signature property of the multi-Weyl
semimetals with $n> 1$. We think that this is indeed a reasonable hypothesis which should be tested carefully
in future experiments. However, we would like to point out that the chemical potentials and/or
temperatures of the order of $\epsilon_0$ probe the band structure sufficiently far from the Weyl nodes.
Therefore, in that region our model predictions may not be very reliable for real materials. This
is not so critical since the topological properties of Weyl semimetals become muted there anyway.

By the same token, we can argue that the lattice model (\ref{model-H-def}) should be reliable (at least
qualitatively) in the region of sufficiently small chemical potentials and temperatures. From the results
shown in Figs.~\ref{fig:Kubo-E-L21-mu} and \ref{fig:Kubo-E-L21-T}, we find that, in the region of small
chemical potentials (temperatures), the multi-Weyl semimetals with $n> 1$ have a much steeper
dependence on $\mu$ ($T$) than their counterparts with the Weyl nodes of the topological charge
$n=1$. In fact, this observation might be rather useful in applications, e.g., when one wants to induce
a large heat flow by applying weak electric fields.

A few words are in order about the off-diagonal coefficient $L_{xy}^{21}$. Its dependencies on the chemical
potential and temperature are shown in Figs.~\ref{fig:Kubo-E-L21-mu}(c) and \ref{fig:Kubo-E-L21-T}(c),
respectively. From a physics viewpoint, this coefficient describes the response in the form of a heat current
perpendicular to the external electric field applied, i.e.,
\begin{equation}
\label{Kubo-Lxy-21-xxx}
\mathbf{J}^{Q}_{\text{{\tiny Ett}}} = \frac{e}{T} L_{xy}^{21} \left[\bm{E}\times \hat{\mathbf{b}}\right],
\end{equation}
where $\hat{\mathbf{b}} \equiv \mathbf{b}/b$. As is easy to check from the analytical expression, the
ratio $L_{xy}^{21}/T$ vanishes in the limit when both the chemical potential and temperature vanish.
In essence, the relation in Eq.~(\ref{Kubo-Lxy-21-xxx}) describes the inverse of the Nernst effect and is sometimes
called the Ettingshausen-Nernst effect. It is clear from our analysis that both effects have topological
roots in the multi-Weyl semimetals. The results in Figs.~\ref{fig:Kubo-E-L21-mu}(c) and \ref{fig:Kubo-E-L21-T}(c)
suggest that the corresponding effect is much more pronounced in the multi-Weyl semimetals with
$n> 1$ than in the Weyl semimetals with $n=1$.

Last but not least let us note that, in view of the Onsager reciprocal relation, $L_{nm}^{21}=L_{nm}^{12}$,
all results obtained in this subsection are also valid for the thermoelectric transport coefficients in the electric current. In
particular, the Nernst conductivity is defined by $L_{xy}^{12}$ and the corresponding current reads
\begin{equation}
\label{Kubo-Lxy-12-xxx}
\mathbf{J}_{\text{{\tiny Ner}}} = e L_{xy}^{21} \left[\bm{\nabla}\left(\frac{1}{T}\right)\times \hat{\mathbf{b}}\right].%
\end{equation}
This is in agreement with the previous findings in Ref.~\cite{Sharma:2016-Weyl}, where the anomalous
Nernst response was predicted for the multi-Weyl semimetals. Because of its explicit dependence on
the chiral shift parameter $\mathbf{b}$, such a contribution would not appear naturally in the conventional chiral
kinetic theory. Thus, in a way, heat and electric currents (\ref{Kubo-Lxy-21-xxx}) and (\ref{Kubo-Lxy-12-xxx}) can be viewed as analogues of the Bardeen--Zumino current. Such a characterization is not rigorous, however, because these currents stem from thermally excited quasiparticles.

\subsection{Coefficient $L_{nm}^{22}$}
\label{sec:Kubo-L22}

Finally, let us calculate the transport coefficient which describes the flow of the heat current in response to
a temperature gradient, i.e., $L_{nm}^{22}=K_{nm}^{22}-2T\epsilon_{nml}M_l^{Q}$, where the corresponding
Kubo's coefficient is defined by Eq.~(\ref{Kubo-L-22-A-1-alt}) and the heat magnetization is given
by Eq.~(\ref{Kubo-magnetization-heat-Streda-int-2}).

After calculating the trace in Eq.~(\ref{Kubo-L-22-A-1-alt}), the expression for the Kubo's coefficient
can be written as a sum of the dissipative and nondissipative terms,
\begin{eqnarray}
\label{Kubo-L-22-A-2-alt}
K_{nm}^{22} = K_{nm}^{22, {\rm ND}} + K_{nm}^{22, {\rm D}},
\end{eqnarray}
where
\begin{eqnarray}
\label{Kubo-L-22-A-2-R-alt}
K_{nm}^{22, {\rm D}} &=& 2\pi\,T \int\frac{d^3\mathbf{k}}{(2\pi)^3} \int d\omega  \frac{(\omega-\mu)^2}{4T\cosh^2{\left(\frac{\omega-\mu}{2T}\right)}}
\sum_{s,s^{\prime}=\pm}\frac{ss^{\prime}}{4|\mathbf{d}|^2}\delta_{\Gamma}\left(\omega-s|\mathbf{d}|\right)\delta_{\Gamma}
\left(\omega-s^{\prime}|\mathbf{d}|\right) \nonumber\\
&\times&\Big[(ss^{\prime}-1) |\mathbf{d}|^2\left(\left(\partial_{k_n}\mathbf{d})\cdot(\partial_{k_m}\mathbf{d}\right)\right)
+2\left(\mathbf{d}\cdot(\partial_{k_n}\mathbf{d})\right)\left(\mathbf{d}\cdot(\partial_{k_m}\mathbf{d})\right) \Big],
\end{eqnarray}
and
\begin{equation}
\label{Kubo-L-22-A-2-I-gamma-not-0}
K_{nm}^{22, {\rm ND}} =
4 T\int\frac{d^3\mathbf{k}}{(2\pi)^3} \int \int d\omega d\omega^{\prime} \frac{\left[n_F(\omega)-n_F(\omega^{\prime})\right](\mu-\omega)
(\mu-\omega^{\prime})}{(\omega-\omega^{\prime})^2}|\mathbf{d}|^2
\delta_{\Gamma}\left(\omega-|\mathbf{d}|\right)\delta_{\Gamma}\left(\omega^{\prime}+|\mathbf{d}|\right)\Omega_{nm}(\mathbf{k}).
\end{equation}
In the clean limit $\Gamma\to0$, the latter reduces to
\begin{equation}
\label{Kubo-L-22-A-2-I-alt}
\lim_{\Gamma\to0}K_{nm}^{22, {\rm ND}}
= T\int\frac{d^3\mathbf{k}}{(2\pi)^3} \left[n_F(|\mathbf{d}|)-n_F(-|\mathbf{d}|)\right] \left(\mu^2-|\mathbf{d}|^2\right)
\Omega_{nm}(\mathbf{k}).
\end{equation}
As in the case of other transport coefficients, after combining the above results for the Kubo's coefficients
with the heat magnetization in Eq.~(\ref{Kubo-magnetization-heat-Streda-int-2}), we find that the only nonzero
components of the heat transport coefficient $L_{nm}^{22}$
%$L_{nm}^{22}=K_{nm}^{22}-2T\epsilon_{nml}M_l^{Q}$
are $L_{xx}^{22}=L_{yy}^{22}$, $L_{zz}^{22}$, and $L_{xy}^{22}=-L_{yx}^{22}$.

\begin{figure}[t]
\begin{center}
\hspace{-0.32\textwidth}(a)\hspace{0.32\textwidth}(b)\hspace{0.32\textwidth}(c)\\[0pt]
\includegraphics[width=0.32\textwidth]{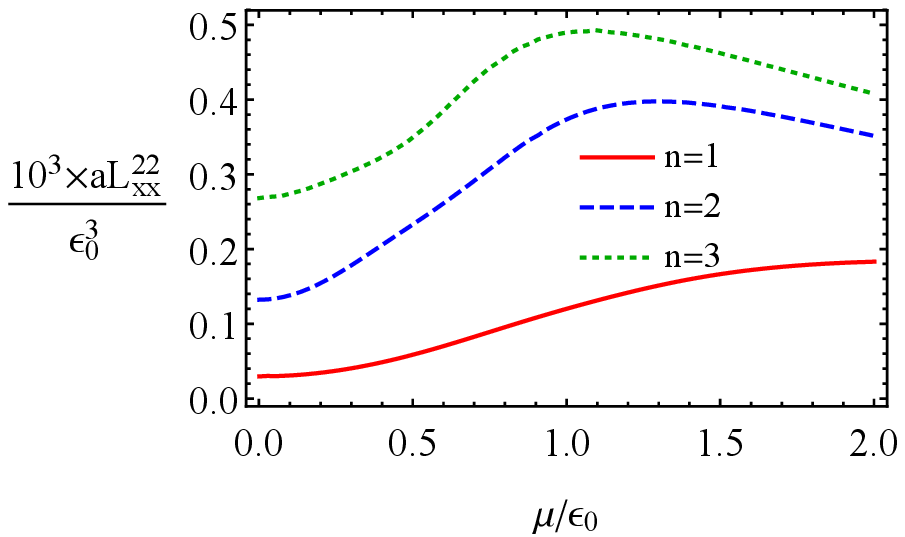}\hfill
\includegraphics[width=0.32\textwidth]{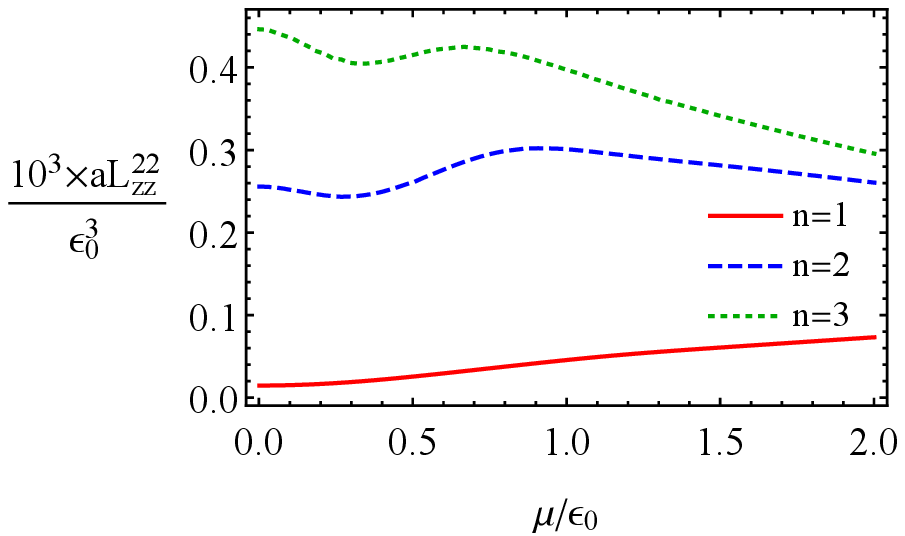}\hfill
\includegraphics[width=0.32\textwidth]{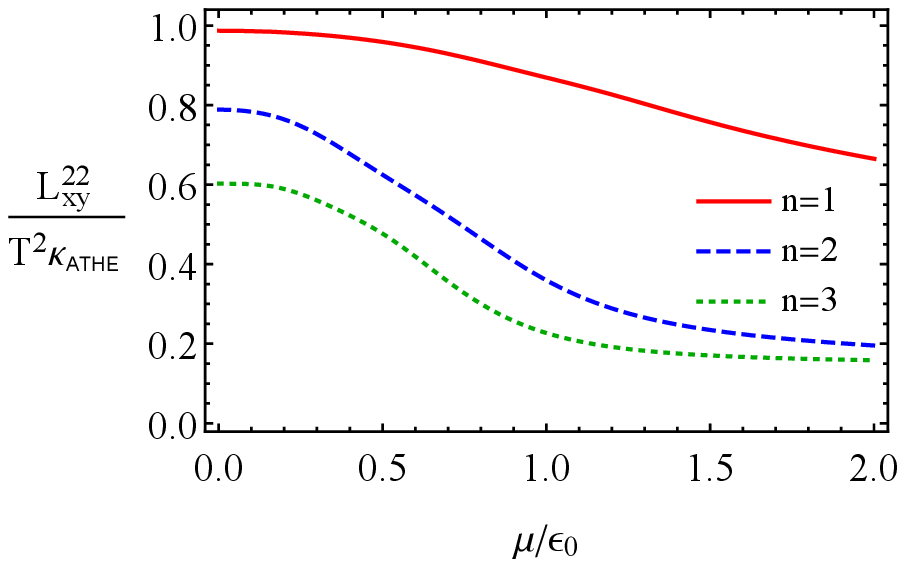}
\caption{The dependence of the transport coefficients $L_{xx}^{22}$, $L_{zz}^{22}$, and $L_{xy}^{22}$ on
the chemical potential in a Weyl semimetal (red solid line), a double-Weyl semimetal (blue dashed line), and
a triple-Weyl semimetal (green dotted line) at fixed $T=0.1\,\epsilon_0$. In panels (a) and (b), the
quasiparticle transport width is modeled by $\Gamma=\Gamma_0\left(1+\omega^2/\epsilon_0^2\right)$ with
$\Gamma_0=0.1\,\epsilon_0$. In panel (c), the results are plotted in the clean limit, $\Gamma=0$.
The numerical values of other model parameters are defined in Appendix~\ref{Sec:App-model}.}
\label{fig:Kubo-E-L22-mu}
\end{center}
\end{figure}

\begin{figure}[t]
\begin{center}
\hspace{-0.32\textwidth}(a)\hspace{0.32\textwidth}(b)\hspace{0.32\textwidth}(c)\\[0pt]
\includegraphics[width=0.32\textwidth]{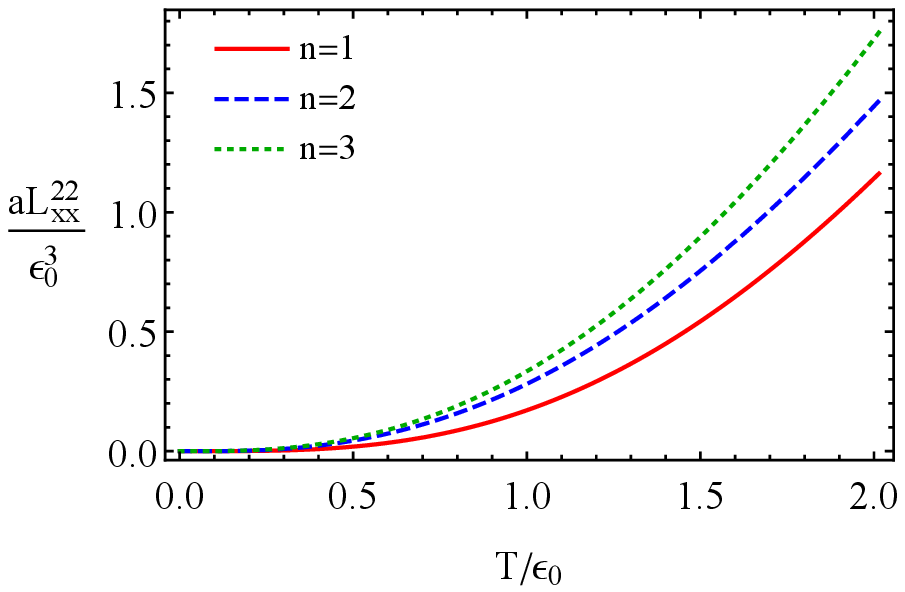}\hfill
\includegraphics[width=0.32\textwidth]{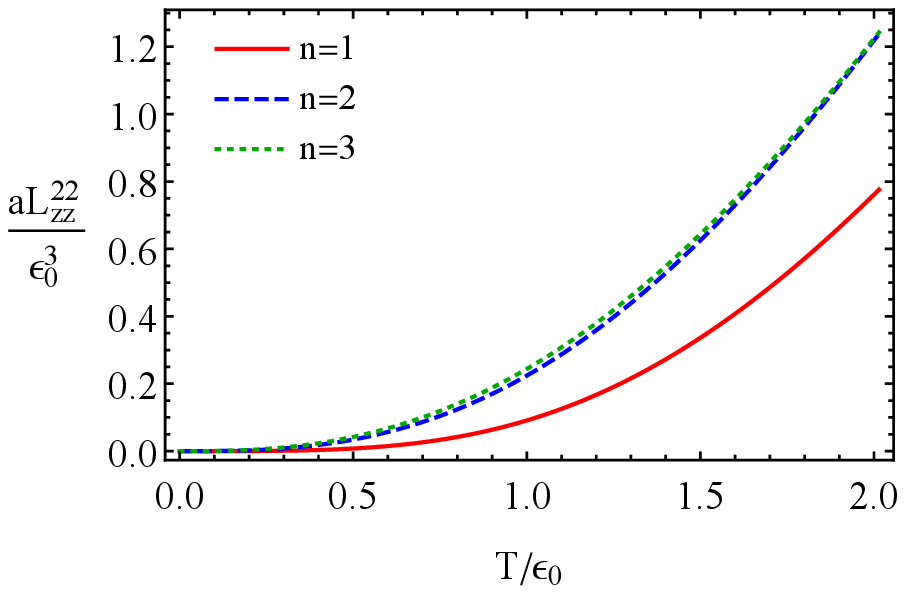}\hfill
\includegraphics[width=0.32\textwidth]{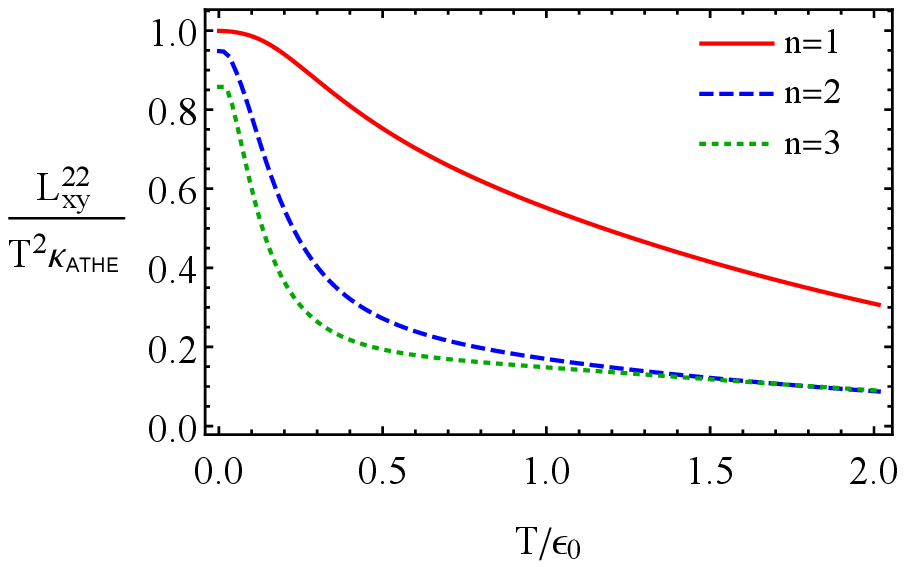}
\caption{The dependence of the transport coefficients $L_{xx}^{22}$, $L_{zz}^{22}$, and $L_{xy}^{22}$ on
temperature in a Weyl semimetal (red solid line), a double-Weyl semimetal (blue dashed line), and
a triple-Weyl semimetal (green dotted line) at fixed $\mu=0.1\,\epsilon_0$. In panels (a) and (b), the
quasiparticle transport width is modeled by $\Gamma=\Gamma_0\left(1+\omega^2/\epsilon_0^2\right)$ with
$\Gamma_0=0.1\,\epsilon_0$. In panel (c), the results are plotted in the clean limit, $\Gamma=0$.
The numerical values of other model parameters are defined in Appendix~\ref{Sec:App-model}.}
\label{fig:Kubo-E-L22-T}
\end{center}
\end{figure}

The numerical results for the coefficients $L_{xx}^{22}$, $L_{zz}^{22}$, and $L_{xy}^{22}$ as functions of
the chemical potential and temperature are shown in Figs.~\ref{fig:Kubo-E-L22-mu} and \ref{fig:Kubo-E-L22-T}.
We used the same model parameters and assumptions as in the calculations of other coefficients in
the previous two subsections.

The results for all multi-Weyl semimetals with topological charges $n=1,2,3$ appear to be qualitatively
similar for each of the three distinct components of the heat transport coefficients $L_{xx}^{22}$, $L_{zz}^{22}$,
and $L_{xy}^{22}$. As in the case of other coefficients, the dependence on the chemical potential
appears to be nonmonotonic for the multi-Weyl semimetals with the topological charge $n>1$, but
not for $n=1$. This is in contrast to the temperature dependence shown in Fig.~\ref{fig:Kubo-E-L22-T},
which is monotonic for all three coefficients $L_{xx}^{22}$, $L_{zz}^{22}$, and $L_{xy}^{22}$.

It should be noted that the off-diagonal coefficient $L_{xy}^{22}$ describes the thermal Hall effect.
In multi-Weyl semimetals, this is also an anomalous effect that is directly related to the topological nature
of the Weyl nodes. In the limit $T\to0$ and $\mu\to0$, as is easy to check from our analytical formulas,
this coefficient coincides with $T^2\kappa_{\text{{\tiny ATHE}}}$, where
\begin{equation}
\kappa_{\text{{\tiny ATHE}}}= -n\frac{Tb_z}{6}
\label{Kubo-E-ATHE}
\end{equation}
is the anomalous thermal Hall conductivity in a multi-Weyl semimetal. In terms of the currents, this
corresponds to
\begin{equation}
\mathbf{J}^{Q}_{\text{{\tiny ATHE}}} = -\frac{n T^3}{6} \left[\bm{\nabla}\left(\frac{1}{T}\right)\times \mathbf{b}\right].
\label{Kubo-E-BZ-thermal}
\end{equation}
As we will see in Sec.~\ref{sec:Kubo-thermo-all}, this anomalous thermal Hall current plays a principal
role in reproducing the Wiedemann-Franz law. Similarly to the Nernst current, this one also depends
explicitly on the chiral shift parameter $\mathbf{b}$ and, thus, may resemble the Bardeen--Zumino
term in the electric current. Strictly speaking, however, such a current is induced by thermally-excited
quasiparticles and, therefore, cannot be rigorously identified as the Bardeen--Zumino current.

%%%%%%%%%%%%%%%%%%
%%%%%%%%%%%%%%%%%%

%%%%%%%%%%%%%%%%%%
%%%%%%%%%%%%%%%%%%

\section{Thermal conductivity, Seebeck tensor, Wiedemann--Franz law, and Mott relation}
\label{sec:Kubo-thermo-all}

By making use of the results obtained in the preceding section, here we will study a range of
physics characteristics (e.g., the thermal conductivity and the Seebeck tensor) in multi-Weyl
semimetals that are relevant for experiment and applications. Furthermore, we test the range
of validity of the Wiedemann-–Franz law and the Mott relation in the Kubo's framework.
Indeed, they hold for a generic system as long as the quasiparticle description of electronic
states remains valid and, consequently, are applicable only in the limit $T\to0$. As expected,
the deviations from these relations will be seen when the temperature is nonzero. In addition,
a finite quasiparticle width $\Gamma$ tends to amplify the deviations.

\subsection{Thermal conductivity and Seebeck tensor}
\label{sec:Kubo-thermo-thermal-cond}

Let us start from the definition of the thermal conductivity tensor $\kappa_{nm}$. It can be
given in terms of the transport coefficients calculated in Sec.~\ref{sec:Kubo-L-All} as follows:
\begin{equation}
\label{Kubo-thermo-all-thermal-cond-def}
\kappa_{nm}=\frac{1}{T^2}\left[L_{nm}^{22} - \frac{1}{T}L_{nl}^{21} (L^{11})^{-1}_{lj} L_{jm}^{12}\right].
\end{equation}
Let us note that the last term in the square brackets comes from enforcing a setup in which
a thermal current is flowing, but there is no electrical one. (For details, see for example
Ref.~\cite{Mahan-book}.)

Before proceeding with the numerical investigations of the thermal conductivity, it is worth
reminding about the approximations that we used in the calculation of the tensor coefficients
$L_{nm}^{\alpha\beta}$. In particular, all dissipative (diagonal) components of the tensors
were calculated by using a phenomenological model of quasiparticles with a small, but
nonzero quasiparticle transport width. This was critical for resolving the otherwise unavoidable singularities in the
expressions for the dissipative terms. At the same time, the nondissipative (off-diagonal)
components of the same tensors were obtained in the clean limit. Of course, this is justifiable
because the nondissipative contributions are of topological origin and remain finite in such
a limit. Moreover, while introducing a small nonvanishing width would considerably
complicate the analysis, the results would not change much anyway. In this section, we
use the same treatment even though the quantities such as the thermal conductivity in
Eq.~(\ref{Kubo-thermo-all-thermal-cond-def}) are defined in terms of mixture of dissipative
and nondissipative components.

\begin{figure}[!ht]
\begin{center}
\hspace{-0.32\textwidth}(a)\hspace{0.32\textwidth}(b)\hspace{0.32\textwidth}(c)\\[0pt]
\includegraphics[width=0.32\textwidth]{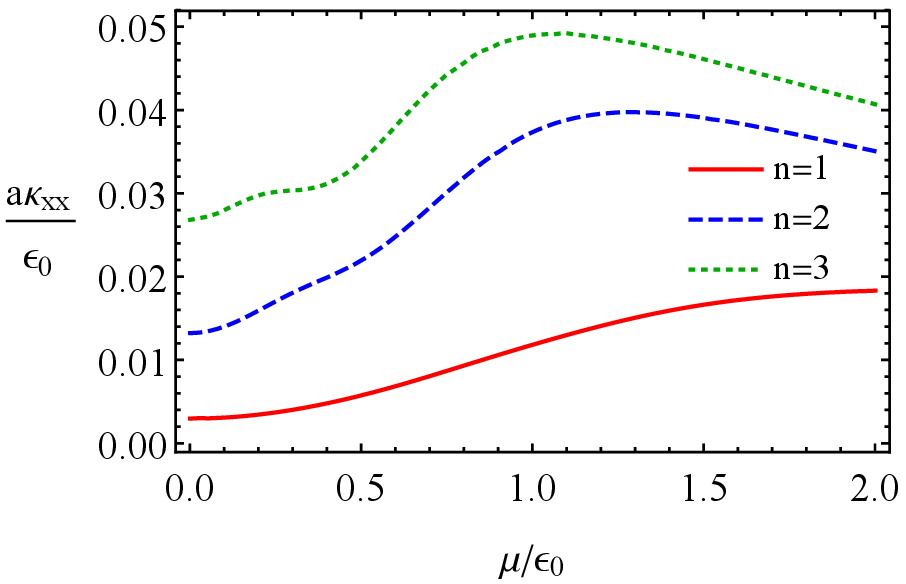}\hfill
\includegraphics[width=0.32\textwidth]{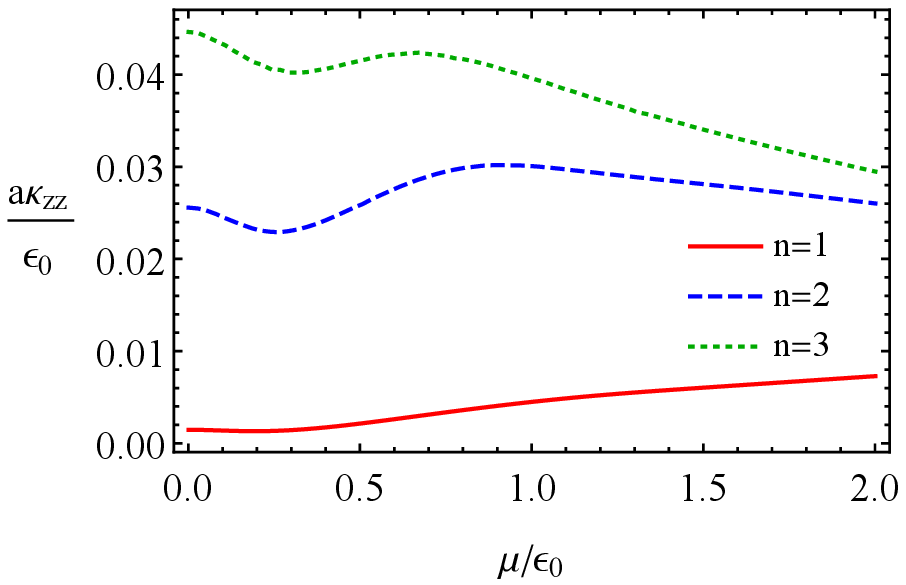}\hfill
\includegraphics[width=0.32\textwidth]{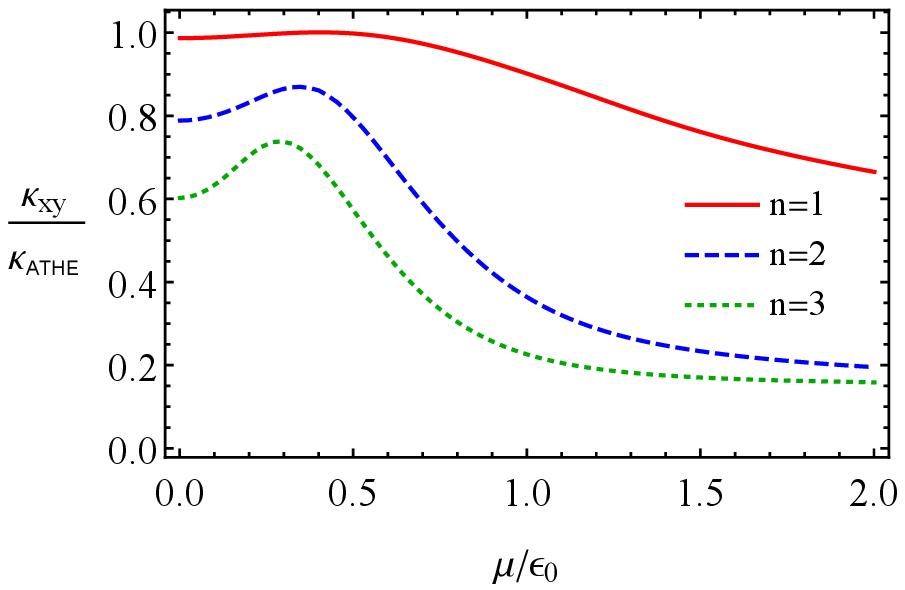}
\caption{The dependence of the thermal conductivity $\kappa_{xx}$, $\kappa_{zz}$, and $\kappa_{xy}$ on
the chemical potential in a Weyl semimetal (red solid line), a double-Weyl semimetal (blue dashed line), a
triple-Weyl semimetal (green dotted line) at fixed $T=0.1\,\epsilon_0$. In panels (a) and (b), the
quasiparticle transport width is modeled by $\Gamma=\Gamma_0\left(1+\omega^2/\epsilon_0^2\right)$ with
$\Gamma_0=0.1\,\epsilon_0$. In panel (c), the results are plotted in the clean limit, $\Gamma=0$.
The numerical values of other model parameters are defined in Appendix~\ref{Sec:App-model}.}
\label{fig:Kubo-E-kappa}
\end{center}
\end{figure}

\begin{figure}[t]
\begin{center}
\hspace{-0.32\textwidth}(a)\hspace{0.32\textwidth}(b)\hspace{0.32\textwidth}(c)\\[0pt]
\includegraphics[width=0.32\textwidth]{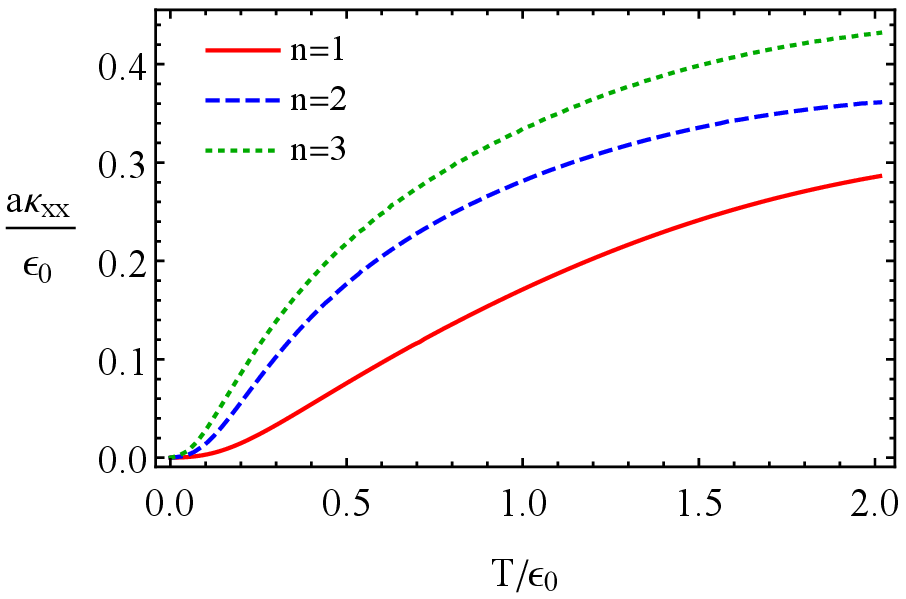}\hfill
\includegraphics[width=0.32\textwidth]{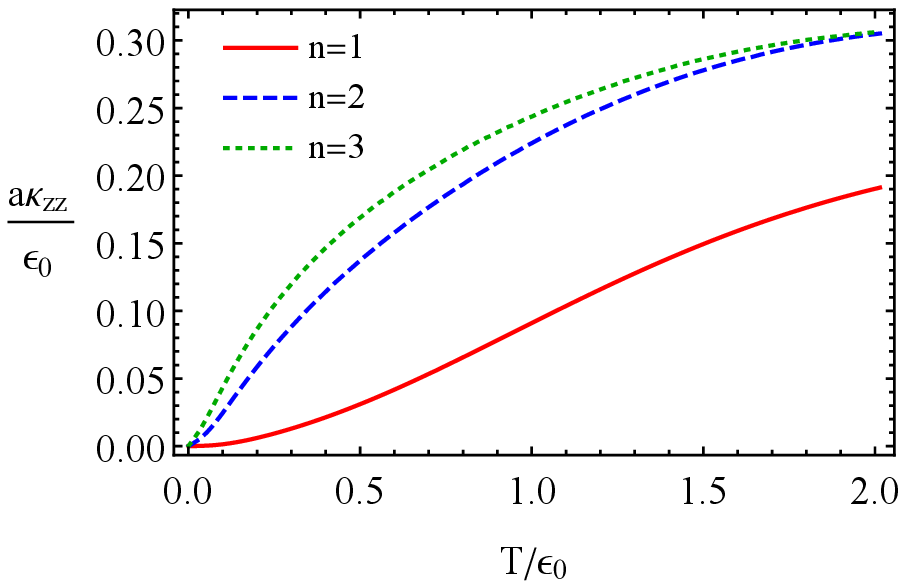}\hfill
\includegraphics[width=0.32\textwidth]{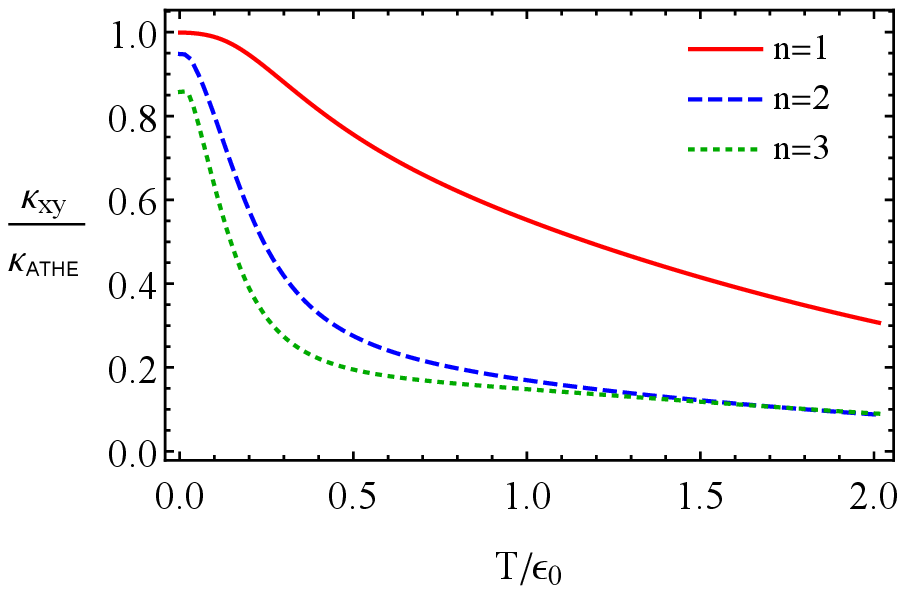}
\caption{The dependence of the thermal conductivity $\kappa_{xx}$, $\kappa_{zz}$, and $\kappa_{xy}$ on
temperature in a Weyl semimetal (red solid line), a double-Weyl semimetal (blue dashed line), a triple-Weyl
semimetal (green dotted line) at fixed $\mu=0.1\,\epsilon_0$. In panels (a) and (b), the quasiparticle
transport width is modeled by $\Gamma=\Gamma_0\left(1+\omega^2/\epsilon_0^2\right)$ with
$\Gamma_0=0.1\,\epsilon_0$. In panel (c), the results are plotted in the clean limit, $\Gamma=0$.
The numerical values of other model parameters are defined in Appendix~\ref{Sec:App-model}.}
\label{fig:Kubo-E-kappa-T}
\end{center}
\end{figure}

We present our numerical results for the three independent components of the thermal
conductivity tensor, i.e., $\kappa_{xx}=\kappa_{yy}$, $\kappa_{zz}$, and $\kappa_{xy}=-\kappa_{yx}$,
in Figs.~\ref{fig:Kubo-E-kappa} and \ref{fig:Kubo-E-kappa-T} as functions of the chemical potential
and temperature, respectively. As is easy to see, the general trends in the dependence of the thermal
conductivity tensor on $\mu$ are rather similar to those of the tensor $L_{nm}^{22}$, shown in
Fig.~\ref{fig:Kubo-E-L22-mu}. %and \ref{fig:Kubo-E-L22-T}.

%%%%%%%%%%%%%%%%%%
%%%%%%%%%%%%%%%%%%

%%%%%%%%%%%%%%%%%%
%%%%%%%%%%%%%%%%%%

As in the case of the electric conductivity, it is instructive to compare the
diagonal components of the thermal conductivity in Eq.~(\ref{Kubo-thermo-all-thermal-cond-def})
with the corresponding results in the linearized chiral kinetic (Boltzmann) theory \cite{Lundgren:2014hra}.
At low temperatures, the latter leads to
\begin{equation}
\kappa_{xx}\propto \frac{\pi^2T}{3\Gamma} \left(\mu^2 + \kappa_0 T^2\right),
%\left(\mu^2+\frac{7\pi^2T^2}{5}\right).
\label{Kubo-thermo-thermal-cond-CKT-kappa}
\end{equation}
where $\kappa_0$ is a numerical coefficient. This dependency qualitatively agrees with the results in Fig.~\ref{fig:Kubo-E-kappa} and at low temperature in Fig.~\ref{fig:Kubo-E-kappa-T}.

\begin{figure}[t]
\begin{center}
\hspace{-0.32\textwidth}(a)\hspace{0.32\textwidth}(b)\hspace{0.32\textwidth}(c)\\[0pt]
\includegraphics[width=0.32\textwidth]{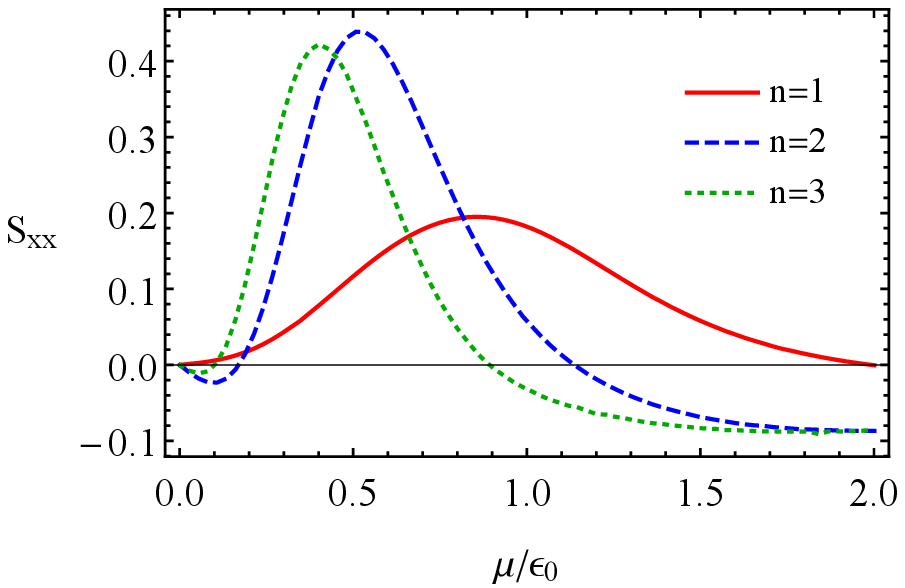}\hfill
\includegraphics[width=0.32\textwidth]{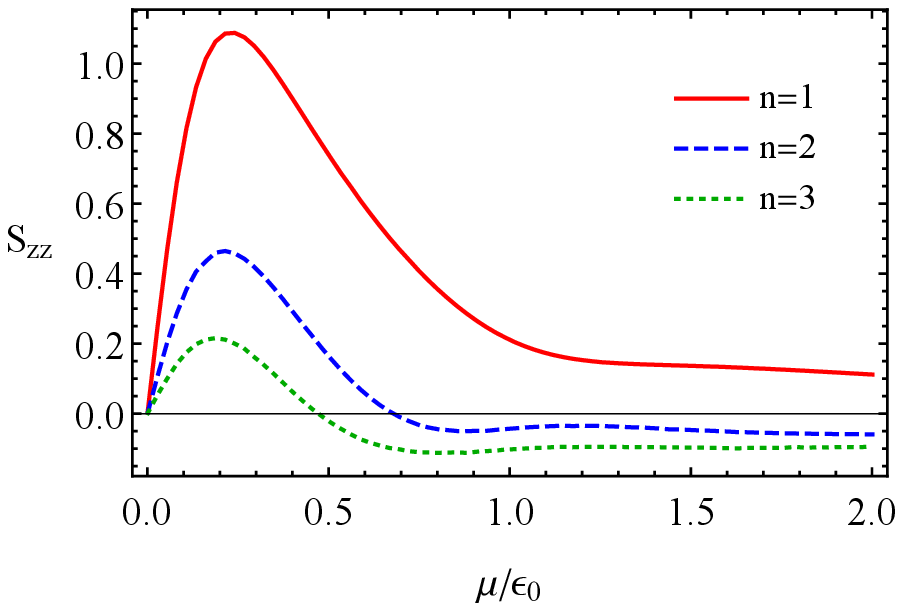}\hfill
\includegraphics[width=0.32\textwidth]{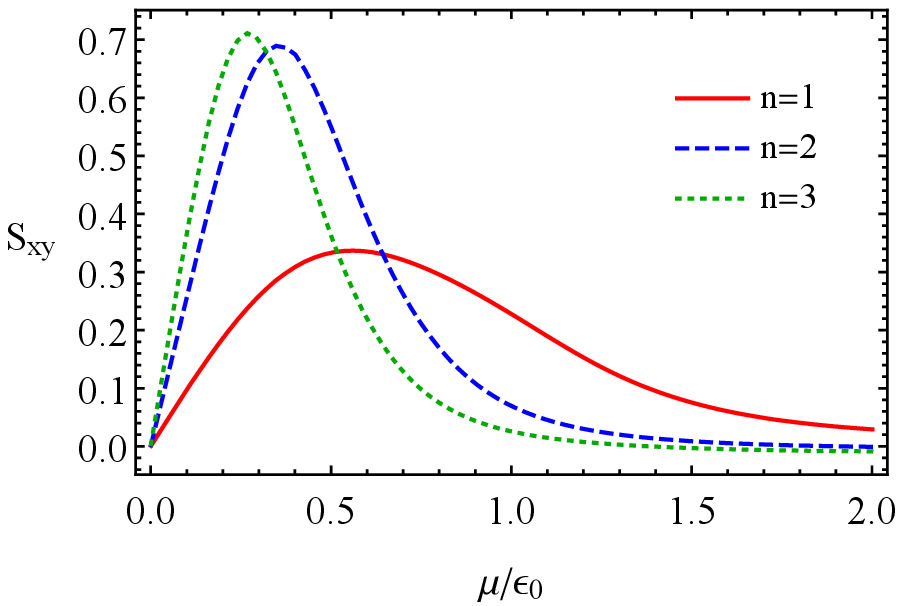}
\caption{The dependence of the thermopower $S_{xx}$, $S_{zz}$, and $S_{xy}$ on
the chemical potential in a Weyl semimetal (red solid line), a double-Weyl semimetal (blue dashed line), a
triple-Weyl semimetal (green dotted line) at fixed $T=0.1\,\epsilon_0$. In panels (a) and (b), the
quasiparticle transport width is modeled by $\Gamma=\Gamma_0\left(1+\omega^2/\epsilon_0^2\right)$ with
$\Gamma_0=0.1\,\epsilon_0$. In panel (c), the results are plotted in the clean limit, $\Gamma=0$.
The numerical values of other model parameters are defined in Appendix~\ref{Sec:App-model}.}
\label{fig:Kubo-E-S}
\end{center}
\end{figure}

\begin{figure}[!ht]
\begin{center}
\hspace{-0.32\textwidth}(a)\hspace{0.32\textwidth}(b)\hspace{0.32\textwidth}(c)\\[0pt]
\includegraphics[width=0.32\textwidth]{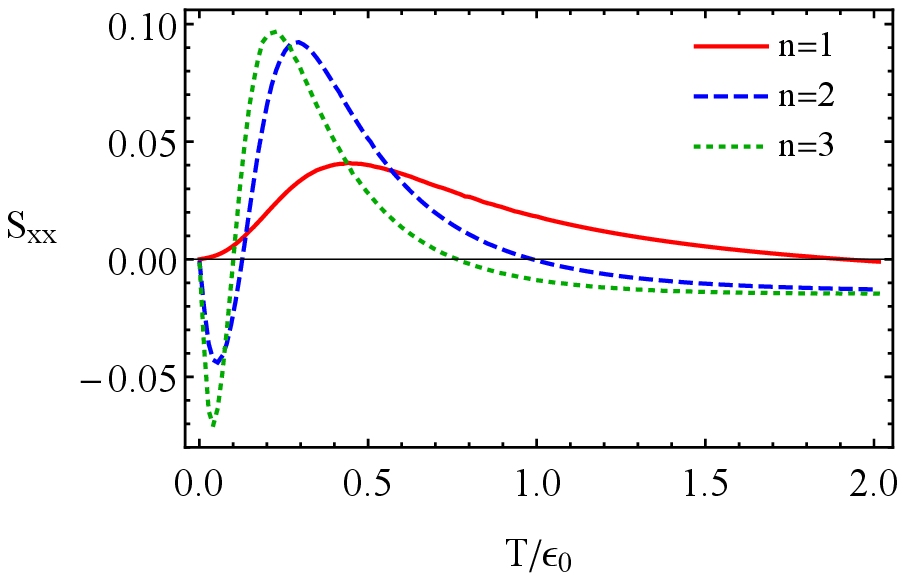}\hfill
\includegraphics[width=0.32\textwidth]{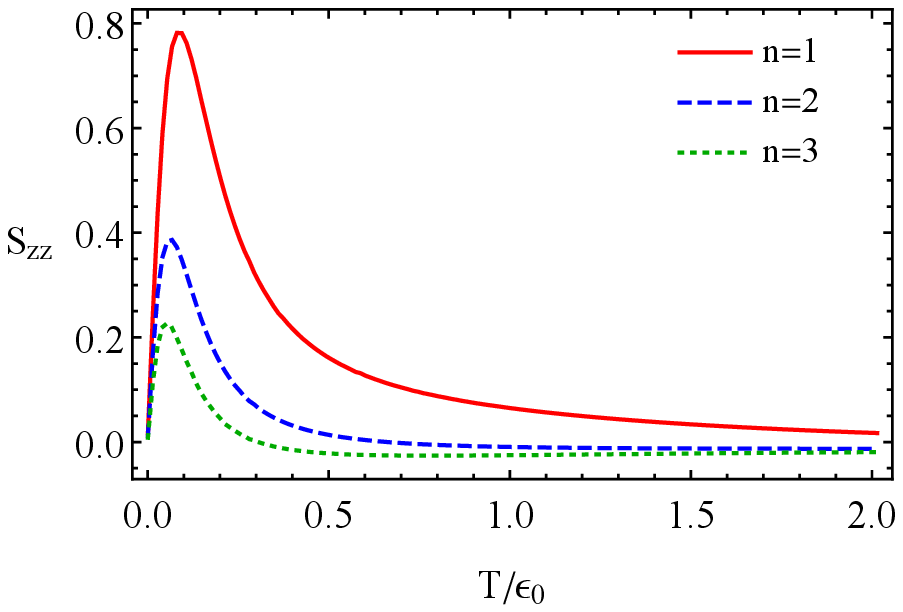}\hfill
\includegraphics[width=0.32\textwidth]{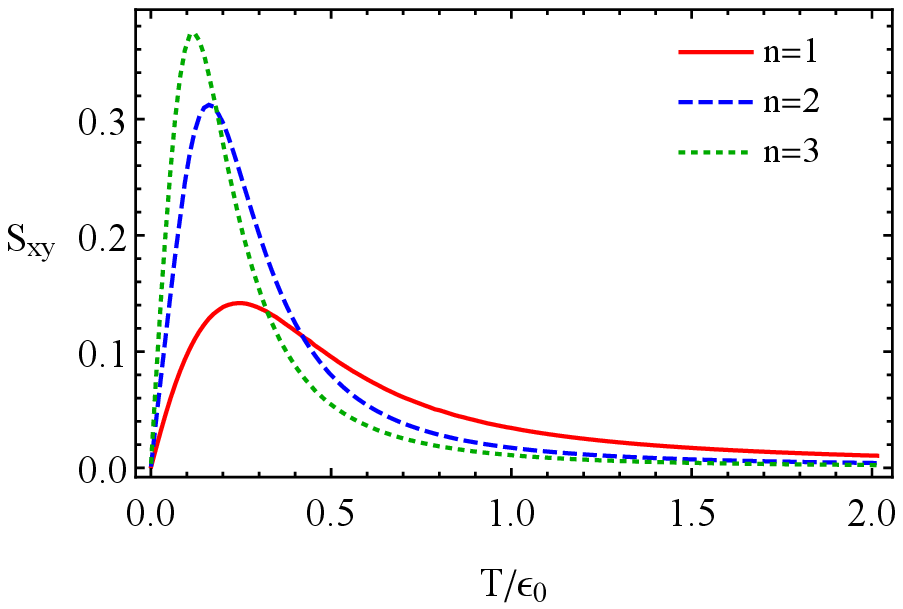}
\caption{The dependence of the thermopower $S_{xx}$, $S_{zz}$, and $S_{xy}$ on
temperature in a Weyl semimetal (red solid line), a double-Weyl semimetal (blue dashed line), a triple-Weyl
semimetal (green dotted line) at fixed $\mu=0.1\,\epsilon_0$. In panels (a) and (b), the quasiparticle
transport width is modeled by $\Gamma=\Gamma_0\left(1+\omega^2/\epsilon_0^2\right)$ with
$\Gamma_0=0.1\,\epsilon_0$. In panel (c), the results are plotted in the clean limit, $\Gamma=0$.
The numerical values of other model parameters are defined in Appendix~\ref{Sec:App-model}.}
\label{fig:Kubo-E-S-T}
\end{center}
\end{figure}

Another important characteristic of the thermal transport is the thermopower, or the Seebeck tensor,
which is defined as
\begin{equation}
\label{Kubo-L-all-Seebeck-def}
S_{nm}=\frac{1}{eT^2}(L^{11})^{-1}_{nl} L_{lm}^{12}.
\end{equation}
We show the dependence of $S_{xx}=S_{yy}$, $S_{zz}$, and $S_{xy}=-S_{yx}$ on the
chemical potential and temperature for multi-Weyl semimetals in Figs.~\ref{fig:Kubo-E-S}
and \ref{fig:Kubo-E-S-T}, respectively. It is interesting to note that the transverse components
of the Seebeck tensor $S_{xx}=S_{yy}$ [see Figs.~\ref{fig:Kubo-E-S}(a)
and \ref{fig:Kubo-E-S-T}(a)] in multi-Weyl semimetals
with $n>1$ have an opposite sign compared to Weyl semimetals with $n=1$ in the region of
small values of $\mu$ or $T$. They also change the sign at relatively large values of $\mu$ or $T$.
We also observe a change of sign for the
longitudinal components of the Seebeck tensor $S_{zz}$, shown in Figs.~\ref{fig:Kubo-E-S}(b)
and \ref{fig:Kubo-E-S-T}(b),
but that change occurs only at relatively large values of the chemical potential $\mu\sim \epsilon_0$ or temperature $T\sim \epsilon_0$.

The common topological feature of both Weyl and multi-Weyl semimetals is a nonzero
off-diagonal component of the Seebeck tensor $S_{xy}$ at nonzero chemical potentials and
temperatures. While all three types of Weyl semimetals share the same bell-shape dependencies
on $\mu$ and $T$, the maximal values of the off-diagonal coefficients are considerably larger
in materials with the topological charge $n>1$.

%%%%%%%%%%%%%%%%%%
%%%%%%%%%%%%%%%%%%

%%%%%%%%%%%%%%%%%%
%%%%%%%%%%%%%%%%%%

\subsection{Wiedemann--Franz law and Mott relation}
\label{sec:Kubo-thermo-WFL}

The Wiedemann--Franz law relates the thermal and electrical conductivities. It is generically
expected to be true when the same well-defined quasiparticles are responsible for both types
of conduction. In this section, we will check the validity of the corresponding law
in our lattice model of multi-Weyl semimetals and study the deviations from it at nonzero
temperature and quasiparticle width.

\begin{figure}[!ht]
\begin{center}
\hspace{-0.32\textwidth}(a)\hspace{0.32\textwidth}(b)\hspace{0.32\textwidth}(c)\\[0pt]
\includegraphics[width=0.32\textwidth]{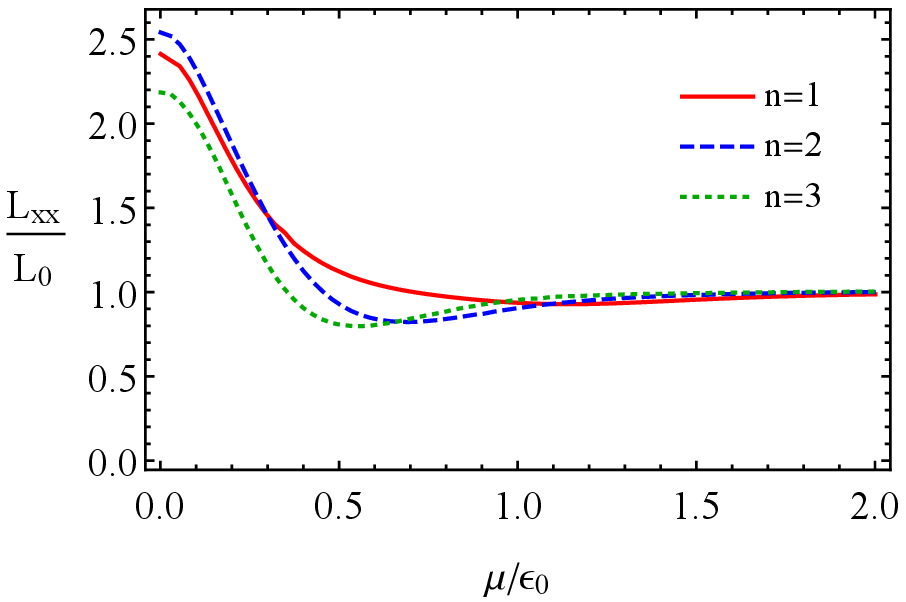}\hfill
\includegraphics[width=0.32\textwidth]{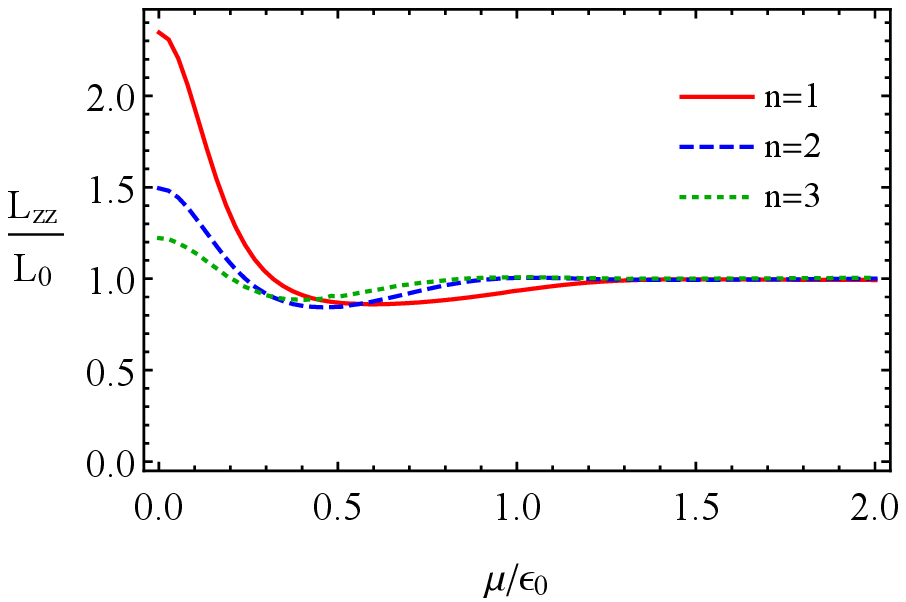}\hfill
\includegraphics[width=0.32\textwidth]{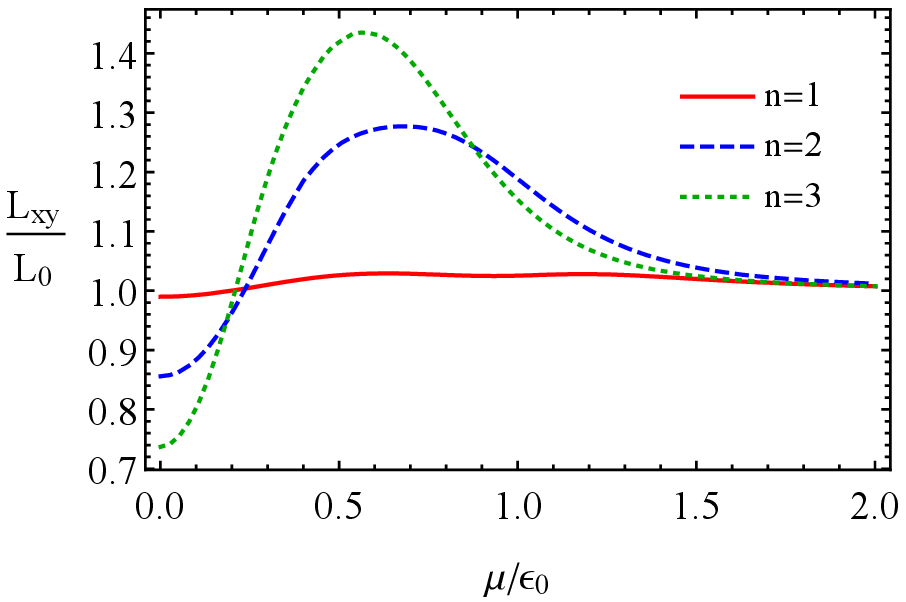}
\caption{The dependence of the relative Lorenz number $L_{xx}/L_0=\kappa_{xx}/(e^2L_0  T L_{xx}^{11})$,
$L_{zz}/L_0=\kappa_{zz}/(e^2L_0  T L_{zz}^{11})$, and $L_{xy}/L_0=\kappa_{xy}/(e^2L_0  T L_{xy}^{11})$ on
the chemical potential in a Weyl semimetal (red solid line), a double-Weyl semimetal (blue dashed line), a
triple-Weyl semimetal (green dotted line) at fixed $T=0.1\,\epsilon_0$. In panels (a) and (b), the
quasiparticle transport width is modeled by $\Gamma=\Gamma_0\left(1+\omega^2/\epsilon_0^2\right)$ with
$\Gamma_0=0.1\,\epsilon_0$. In panel (c), the results are plotted in the clean limit, $\Gamma=0$.
The numerical values of other model parameters are defined in Appendix~\ref{Sec:App-model}.}
\label{fig:Kubo-E-L}
\end{center}
\end{figure}

\begin{figure}[!ht]
\begin{center}
\hspace{-0.32\textwidth}(a)\hspace{0.32\textwidth}(b)\hspace{0.32\textwidth}(c)\\[0pt]
\includegraphics[width=0.32\textwidth]{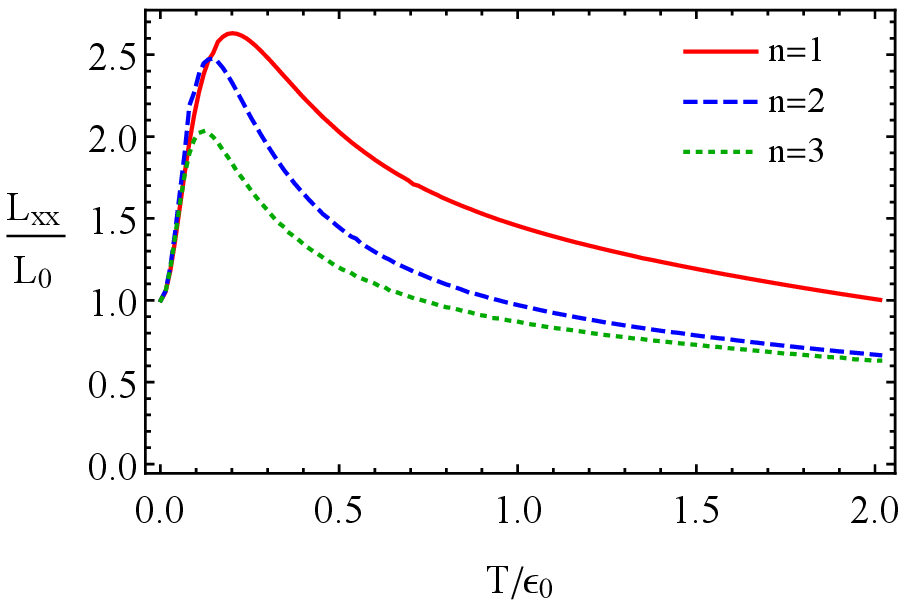}\hfill
\includegraphics[width=0.32\textwidth]{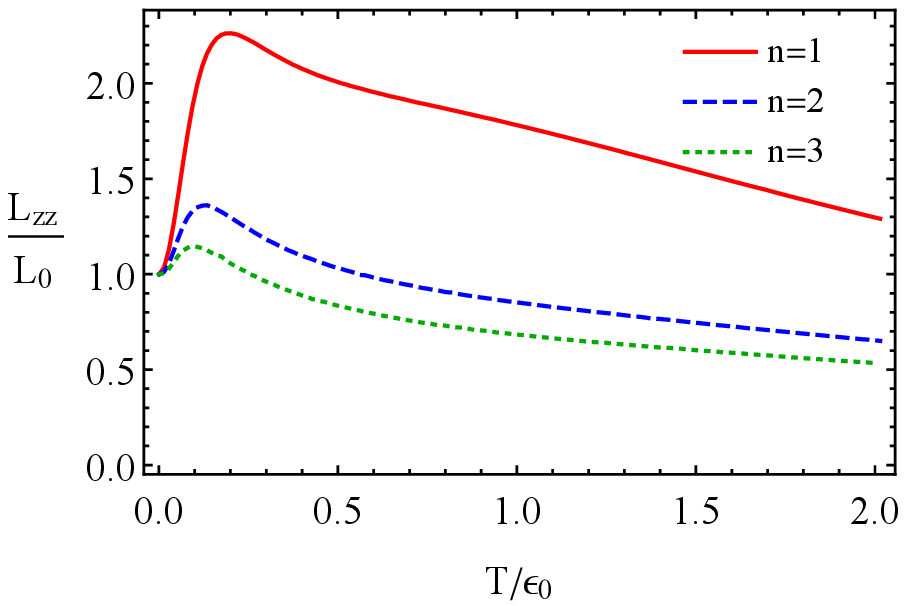}\hfill
\includegraphics[width=0.32\textwidth]{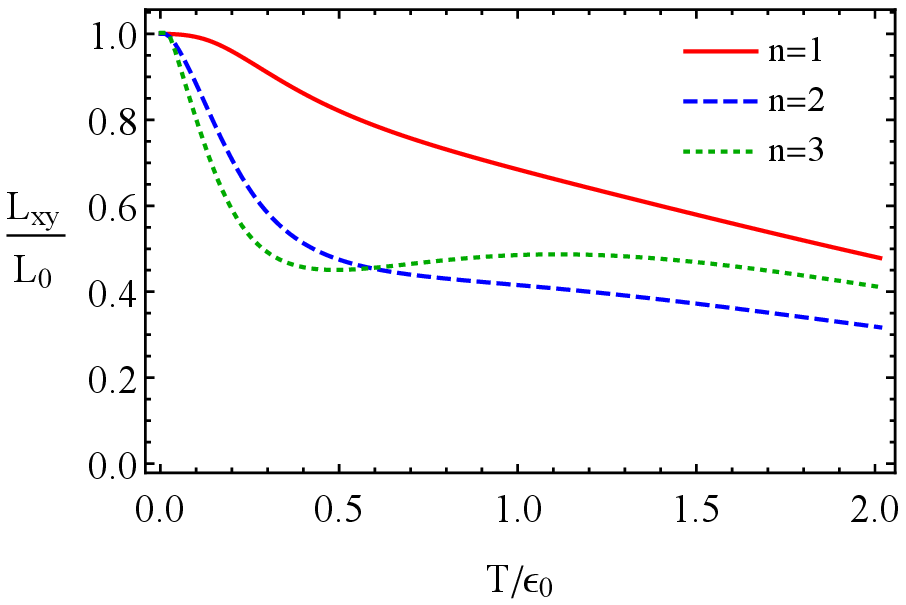}
\caption{The dependence of the relative Lorenz number $L_{xx}/L_0=\kappa_{xx}/(e^2L_0  T L_{xx}^{11})$,
$L_{zz}/L_0=\kappa_{zz}/(e^2L_0  T L_{zz}^{11})$, and $L_{xy}/L_0=\kappa_{xy}/(e^2L_0  T L_{xy}^{11})$ on
the chemical potential in a Weyl semimetal (red solid line), a double-Weyl semimetal (blue dashed line), a
triple-Weyl semimetal (green dotted line) at fixed $\mu=0.1\,\epsilon_0$. In panels (a) and (b), the
quasiparticle transport width is modeled by $\Gamma=\Gamma_0\left(1+\omega^2/\epsilon_0^2\right)$ with
$\Gamma_0=0.1\,\epsilon_0$. In panel (c), the results are plotted in the clean limit, $\Gamma=0$.
The numerical values of other model parameters are defined in Appendix~\ref{Sec:App-model}.}
\label{fig:Kubo-E-L-T}
\end{center}
\end{figure}

In terms of the transport coefficients, the Wiedemann--Franz law reads
\begin{equation}
\label{Kubo-L-all-WFL-def}
\kappa_{nm}= e^2 L_0  T  L_{nm}^{11},
\end{equation}
where $L_0=\pi^2/(3e^2)$ denotes the Lorenz number.
In order to study this relation in multi-Weyl semimetals, we plot the dependence of each of the three independent
components of the relative Lorenz number $L_{nm}/L_0\equiv \kappa_{nm}/(e^2 L_0  T L_{nm}^{11})$
on the chemical potential and temperature in Figs.~\ref{fig:Kubo-E-L} and \ref{fig:Kubo-E-L-T},
respectively.

As we see from Figs.~\ref{fig:Kubo-E-L}(a) and \ref{fig:Kubo-E-L}(b), there are substantial
deviations from the naive behavior predicted by the Wiedemann--Franz law in the transverse
and longitudinal components, $L_{xx}$ and $L_{zz}$, when $\mu$ is small. This is due to the fact that
the quasiparticle description breaks down when $T\gtrsim \mu$. As a careful analysis shows,
the effect of nonzero temperature is further amplified by a nonvanishing quasiparticle width $\Gamma$.
Overall, the dependencies of all relative Lorenz number components
are qualitatively similar in a Weyl semimetal and its multi-Weyl counterparts.
However, this is not the case for the off-diagonal components of the relative Lorenz number. The
latter are quite different for multi-Weyl semimetals with different topological charges.

From the temperature dependence in Fig.~\ref{fig:Kubo-E-L-T}, we see that, as expected, the
Wiedemann--Franz law holds in the limit of small $T$. As for the deviations at nonzero $T$, they
first quickly increase with temperature and then gradually decrease. In the case of the relative
Lorenz numbers $L_{xx}/L_0$ and $L_{zz}/L_0$, the deviations in the intermediate region of
temperatures are larger in the $n=1$ Weyl semimetal than in the double- and triple-Weyl semimetals.
As is clear from Fig.~\ref{fig:Kubo-E-L-T}(c), however, the situation is opposite for
$L_{xy}/L_0$.

Here, it is important to emphasize that the Wiedemann--Franz law holds exactly in the limit $T \to 0$
and $\Gamma\to 0$. For the details of the corresponding analysis, see Appendix~\ref{sec:App-WFL-T0}.
This result clearly demonstrates that a nontrivial topology in the multi-Weyl semimetals by itself does not
cause any violation of the Wiedemann--Franz law. This also agrees with the analysis in the linearized kinetic
theory \cite{Lundgren:2014hra}.

%%%%%%%%%%%%%%%%%%
%%%%%%%%%%%%%%%%%%

%%%%%%%%%%%%%%%%%%
%%%%%%%%%%%%%%%%%%

Let us finally discuss the Mott relation, i.e.,
\begin{equation}
\label{Kubo-L-all-Mott-def}
L_{nm}^{12}= \frac{\pi^2 T^3}{3}\frac{d L_{nm}^{11} }{d\mu},
\end{equation}
which is expected to hold at low temperature. Similarly to the case of the Wiedemann--Franz law,
we find that small deviations from the Mott relation show up only with an increase of temperature,
when the quasiparticle description of electronic states starts to gradually break down.

\section{Summary and Discussions}
\label{sec:Summary-Discussions}

In this paper, by using a generic lattice model, we studied the thermoelectric properties
of multi-Weyl semimetals with a broken time-reversal symmetry. The calculations are performed
in the Kubo's linear response theory
that take into account the additional contributions connected with the electromagnetic orbital
and heat magnetizations. These contributions appear due to the modification of the charge and
heat current operators in the Luttinger method, where a gravitational field is introduced as
the mechanical counterpart of the temperature gradient. While these magnetizations do not affect
the non-anomalous diagonal thermoelectric transport coefficients, their presence is absolutely
crucial in the anomalous ones and guarantees the validity of the Wiedemann--Franz law and the Mott relation.

As in the case of the electric response studied previously by us using the same lattice model in
Ref.~\cite{Gorbar:2017-Bardeen}, the nontrivial topology of the electron structure of multi-Weyl
semimetals also plays a profound role in the thermoelectric transport. Indeed, the topological charge
of the Weyl nodes causes the anomalous Nernst effect, which implies the existence of an electric
current in response to a thermal gradient in the absence of an external magnetic field. Similarly,
the off-diagonal components of the heat current are induced by a thermal gradient and an
electric field. They describe the anomalous thermal Hall and Nernst effects, respectively.
In this connection, it should be noted that these anomalous effects could not be correctly reproduced
in the linearized chiral kinetic theory, unless the latter is supplemented by the Bardeen--Zumino
currents \cite{Gorbar:2016ygi} or the appropriate Berry curvature from a lattice model \cite{Sharma:2016-Weyl}. This is in contrast to the nonanomalous response coefficients which are qualitatively
the same in all frameworks including the chiral kinetic (Boltzmann) theory with a linear dispersion law.

Our calculations show that all anomalous thermoelectric coefficients in multi-Weyl semimetals
contain an additional multiplication factor, which in the limit of zero temperature and chemical potential
is the integer topological charge of the Weyl nodes. This conclusion also agrees with the previously
obtained results in Ref.~\cite{Chen-Fiete:2016}, where the double-Weyl model was studied, as well
as with the analysis in Ref.~\cite{1705.04576}, where the high-energy-inspired Fujikawa method
was employed. We would like to mention also that the topological contribution to the thermal current
takes a form which is somewhat similar to the electromagnetic Bardeen--Zumino current
\cite{Landsteiner:2013sja,Landsteiner:2016,Gorbar:2017-Bardeen}. However, it is a current
induced by thermally excited quasiparticles and, thus, not a true analog of the Bardeen--Zumino current.

In this paper, we studied in detail the dependence of the thermoelectric coefficients in multi-Weyl semimetals
(with the topological charges of Weyl nodes $n=1,2,3$) on the chemical potential and temperature. In general,
we found that the corresponding dependence is much milder in the $n=1$ Weyl semimetals, compared to the
double- and triple-Weyl materials. Also, as one might expect in the case of the larger topological charges, the
anomalous response is much more susceptible to the chemical potential and temperature when $n> 1$.
This is the case for the anomalous Hall, Ettingshausen-Nernst, Nernst, and thermal Hall effects.

Interestingly, we found that the diagonal components of the Seebeck tensor in the double- and triple-Weyl
semimetals can change the sign as functions of $\mu$ and $T$. However, this property is not shared by the
$n=1$ Weyl semimetals. It may be also important to mention that the non-topological diagonal thermoelectric
coefficients are typically several times larger for multi-Weyl semimetals than for the Weyl semimetals with
$n=1$. One might speculate, therefore, that the multi-Weyl semimetals may be more promising for application
in thermoelectric devices.

Within the Kubo's formalism, we checked that the results for the thermoelectric coefficients
in multi-Weyl semimetals agree with the Wiedemann-Franz law and the Mott relation in the limit of
zero temperature. We also found that deviations appear and grow with increasing values of
temperature and are further amplified by a quasiparticle width. As is clear, such deviations
indicate that the quasiparticle description of the electronic states starts to gradually fail, which is
indeed expected when $T\gtrsim \mu$ and $\Gamma\gtrsim\mu$. (This finding is also in agreement
with the results in Ref.~\cite{Tabert:2016}.)

Last but not least, let us briefly discuss the relevance of the obtained results for Weyl
semimetals with a broken inversion, but intact TR symmetry. In such materials the total number
of Weyl points should be a multiple of four (see, e.g., Ref.~\cite{Armitage-Vishwanath:2017-Rev}).
This is the consequence of the time-reversal symmetry that maps each pair of opposite-chirality Weyl
nodes separated by $2\mathbf{b}$ in momentum space to another pair of Weyl nodes separated by
$-2\mathbf{b}$. Clearly, for such Weyl semimetals, the sum of all chiral shifts must vanish,
i.e., $\sum_{n} \mathbf{b}^{(n)} =\mathbf{0}$. Then, since all anomalous thermoelectric responses,
i.e., the anomalous Hall, Nernst, Ettingshausen-Nernst, and thermal Hall conductivities, are linear
in the chiral shift vector, we can expect that such effects are absent in the inversion symmetry
broken Weyl semimetals (which is in agreement with Ref.~\cite{Lundgren:2014hra}).
This would not apply, however, to Weyl semimetals, in which both the inversion and time-reversal
symmetries are broken. The anomalous response in such a general case may be similar
to that in the Weyl semimetals with a broken TR symmetry, but with the chiral shift replaced by
$\mathbf{b}_{\rm eff}\equiv \sum_{n} \mathbf{b}^{(n)} \neq \mathbf{0}$. However, the study
of Weyl semimetals with a broken inversion symmetry clearly deserves a further in-depth investigation,
which is beyond the scope of this study.

\begin{acknowledgments}
The work of E.V.G. was partially supported by the Program of Fundamental Research of the
Physics and Astronomy Division of the National Academy of Sciences of Ukraine.
The work of V.A.M. and P.O.S. was supported by the Natural Sciences and Engineering Research Council of Canada.
The work of I.A.S. was supported by the U.S. National Science Foundation under Grants PHY-1404232
and PHY-1713950.
\end{acknowledgments}

\appendix

\section{Model parameters}
\label{Sec:App-model}

In this appendix, we present a representative set of model parameters which we employ in our
numerical calculations throughout the paper. In order to have a realistic model, will use the
parameters for $\mathrm{Na_3Bi}$ presented in Ref.~\cite{Wang}. The parametrization in
the given paper is related to the notations in Eq.~(\ref{model-H-def}) as follows:
\begin{eqnarray}
\label{lattice-coeff-C-be}
&&t_0=M_0-t_1-2t_2, \qquad t_{1,2}=-\frac{2M_{1,2}}{a^2},\\
&&g_0=C_0-g_1-2g_2, \qquad g_{1,2}=-\frac{2C_{1,2}}{a^2},\\
&&\Lambda=\frac{A}{a},
\label{lattice-coeff-C-ee}
\end{eqnarray}
where
\begin{equation}
\begin{array}{lll}
 C_0 = -0.06382~\mbox{eV},\qquad
& C_1 = 8.7536~\mbox{eV\,\AA}^2,\qquad
& C_2 = -8.4008~\mbox{eV\,\AA}^2,\\
 M_0=0.08686~\mbox{eV},\quad
& M_1=-10.6424~\mbox{eV\,\AA}^2,\qquad
& M_2=-10.3610~\mbox{eV\,\AA}^2,\\
 A=2.4598~\mbox{eV\,\AA}.
\end{array}
\label{lattice-model-parameters}
\end{equation}
For the sake of simplicity, we assume that the Weyl semimetal model has a cubic
lattice, i.e., $a_x=a_y=a_z=a=7.5~\mbox{\AA}$. Although usually this is not the case in real
materials, such an assumption has no effect on the validity of the main qualitative
results in our study.

\section{Matsubara sums}
\label{Sec:App-Matsubara}

In this appendix, we present the results for several types of Matsubara sums needed in the calculation
of the current-current correlators in the main text. By omitting the standard derivation steps, here we
quote only the final results for the following three types of sums:
\begin{eqnarray}
\label{Kubo-L-sum-calc-1}
&& T \sum_{l=-\infty}^{\infty}
\frac{1}{\left(i\omega_l+\mu-\omega\right)\left(i\omega_l-i\Omega_r+\mu-\omega^{\prime}\right)}
=\frac{n_F(\omega)-n_F(\omega^{\prime})}{\omega-\omega^{\prime}-\Omega-i0},\\
\label{Kubo-L-sum-calc-2}
&& T \sum_{l=-\infty}^{\infty}
\frac{i\omega_l}{\left(i\omega_l+\mu-\omega\right)\left(i\omega_l-i\Omega_r+\mu-\omega^{\prime}\right)} = \frac{(\omega-\mu)n_F(\omega)-(\omega^{\prime}-\mu+\Omega)n_F(\omega^{\prime})}{\omega-\omega^{\prime}-\Omega-i0},\\
\label{Kubo-L-sum-calc-3}
&& T \sum_{l=-\infty}^{\infty}
\frac{i\omega_l(i\omega_l-i\Omega_r)}{\left(i\omega_l+\mu-\omega\right)\left(i\omega_l-i\Omega_r+\mu-\omega^{\prime}\right)} =\frac{(\omega-\mu)(\omega-\mu-\Omega)n_F(\omega)-(\omega^{\prime}-\mu+\Omega)(\omega^{\prime}-\mu)n_F(\omega^{\prime})}{\omega-\omega^{\prime}-\Omega-i0},
\end{eqnarray}
where $n_{F}(\omega)=1/\left[e^{(\omega-\mu)/T}+1\right]$ is the Fermi-Dirac distribution function. Note that
$\Omega_r=2\pi r T$ with $r\in \mathbb{Z}$ is the {\em bosonic} Matsubara frequency that corresponds to the
external line in the current-current correlator. When making the analytic continuation to the real axis in the
complex frequency plane, we replaced $i\Omega_r \rightarrow \Omega+i0$. It should be also noted that,
because of the divergent sum in Eq.~(\ref{Kubo-L-sum-calc-3}), the corresponding final result is defined up to an
infinite constant. However, as was shown in Ref.~\cite{Griffin} (see, also, Appendix~B in Ref.~\cite{Sharapov:2003}),
this divergence stems from an improper treatment of time derivatives inside the time-ordered product of the
heat currents. The divergence disappears when the problem is treated more carefully. Thus, the correct
prescription is to ignore the divergent constant term.

\section{The Wiedemann--Franz law at small temperatures and vanishing chemical potential}
\label{sec:App-WFL-T0}

As we saw from the numerical analysis of the thermoelectric transport in Sec.~\ref{sec:Kubo-thermo-WFL},
there are clear deviations from the Wiedemann--Franz at nonzero temperature $T$. Such deviations indicate
that the quasiparticle description starts to fail gradually with increasing $T$ that is further amplified by a quasiparticle
width $\Gamma$. Here we demonstrate analytically that the Wiedemann--Franz law is valid for multi-Weyl
semimetals in the clean limit when $T\to0$.

By setting $\mu=0$ and considering the limit of small temperatures, we derive the following expressions
for the off-diagonal nondissipative components (which should be the most sensitive to the nontrivial topology) of the transport coefficients $K_{nm}^{11}$, $K_{nm}^{21}$,
and  $K_{nm}^{22}$ (with $n\neq m$):
\begin{eqnarray}
\label{Kubo-L-11-WFL}
K_{nm}^{11} &=&-\int\frac{d^3 \mathbf{k}}{(2\pi)^3} \tanh{\left(\frac{|\mathbf{d}|}{2T}\right)}
\Omega_{nm}(\mathbf{k}) \simeq -\int\frac{d^3 \mathbf{k}}{(2\pi)^3} \Omega_{nm}(\mathbf{k}) ,\\
\label{Kubo-L-21-WFL}
K_{nm}^{21} &=&K_{nm}^{12}= \epsilon_{nml}M_l=0,\\
\label{Kubo-L-22-WFL}
K_{nm}^{22} &=& T\int\frac{d^3 \mathbf{k}}{(2\pi)^3} \tanh{\left(\frac{|\mathbf{d}|}{2T}\right)} |\mathbf{d}|^2
\Omega_{nm}(\mathbf{k}) \simeq T
\int\frac{d^3 \mathbf{k}}{(2\pi)^3} |\mathbf{d}|^2 \Omega_{nm}(\mathbf{k}),
\end{eqnarray}
which follow from the more general representations in Eqs.~(\ref{Kubo-L-11-A-2-I}), (\ref{Kubo-L-21-A-2-I-alt}),
and (\ref{Kubo-L-22-A-2-I-alt}), respectively. [Note that all dissipative contributions vanish after the integration
over the whole Brillouin zone.]

In the same small temperature limit, the heat magnetization (\ref{Kubo-magnetization-heat-Streda-int-2})
is given by
\begin{equation}
\label{Kubo-magnetization-heat-Streda-WFL}
M_l^Q  \simeq \frac{\epsilon_{nml}}{2}\int \frac{d^3 \mathbf{k}}{(2\pi)^3}
\frac{\Omega_{nm}(\mathbf{k})}{2}\left(|\mathbf{d}|^2+\frac{\pi^2T^2}{3}\right) .
\end{equation}
By combining all these results, we derive the following expression for the off-diagonal components of the
heat conductivity:
\begin{equation}
\kappa_{nm}=\frac{L_{mn}^{22}}{T^2}= \frac{K_{mn}^{22}-2T\epsilon_{nml}M_l^Q}{T^2}
\simeq -\frac{\pi^2 T }{3}\int\frac{d^3\mathbf{k}}{(2\pi)^3} \Omega_{nm}(\mathbf{k}).
\end{equation}
Now, by taking into account that $L_{nm}^{11}=K_{nm}^{11}$ and using the result in Eq.~(\ref{Kubo-L-11-WFL}),
we find that the Wiedemann--Franz law in Eq.~(\ref{Kubo-L-all-WFL-def}) is not violated or modified
by the nontrivial topology in the multi-Weyl semimetals.

\end{document}